\documentclass[fleqn,usenatbib,useAMS]{mnras}

\usepackage[T1]{fontenc}

\usepackage{graphicx}	
\usepackage{amsmath}	
\usepackage{amssymb}	
\usepackage{wasysym}
\usepackage{makecell}
\usepackage{multicol}  
\usepackage{bm}		
\usepackage{pdflscape}	

\title[Dissolution of tidally filling star clusters harbouring black hole subsystems]{\textsc{Mocca-Survey Database I}: Dissolution of tidally filling star clusters harbouring black hole subsystems}

\author[Giersz et al.]{
M. Giersz,$^{1}$\thanks{E-mail: mig@cmk.edu.pl (MG)}
A. Askar,$^{2}$
L. Wang,$^{3,4,5}$
A. Hypki,$^{6}$
A. Leveque$^{1}$
R. Spurzem$^{7,8,9}$\thanks{RS: Research Fellow at Frankfurt Institute for Advanced Studies}
\\
$^{1}$Nicolaus Copernicus Astronomical Center, Polish Academy of Sciences, ul. Bartycka 18, Warsaw 00-716 Poland\\
$^{2}$Lund Observatory, Department of Astronomy, and Theoretical Physics, Lund University, Box 43, SE-221 00 Lund, Sweden\\
$^{3}$ Helmholtz-Institut f{\"u}r Strahlen- und Kernphysik, University of Bonn, Nussallee 14-16, D-53115 Bonn, Germany\\
$^{4}$Argelander Institut F{\"u}r Astronomie, Auf Dem H{\"u}gel 71, 53121, Bonn, Germany\\
$^{5}$ RIKEN Center for Computational Science, 7-1-26 Minatojima-minami-machi, Chuo-ku, Kobe, Hyogo 650-0047, Japan\\
$^{6}$Astronomical Observatory Institute, Faculty of Physics, A. Mickiewicz University, S\l{}oneczna 36, 60-286 Pozna\'n, Poland\\
$^{7}$National Astronomical Observatories and Key Laboratory for Computational Astrophysics, Chinese Academy of Sciences, \\  20A Datun Road, Chaoyang District, Beijing 100012, China\\
$^{8}$Astronomisches Rechen-Institut, Zentrum f{\"u}r Astronomie, University of Heidelberg, M¨onchhofstrasse 12-14, \\  D-69120 Heidelberg, Germany\\
$^{9}$Kavli Institute for Astronomy and Astrophysics, Peking University, Yi He Yuan Lu 5, HaiDian District, Beijing 100871, China
}

\date{Accepted XXX. Received YYY; in original form ZZZ}

\pubyear{2018}

\begin{document}
\label{firstpage}
\pagerange{\pageref{firstpage}--\pageref{lastpage}}
\maketitle

\begin{abstract}
We investigate the dissolution process for dynamically evolving star
clusters embedded in an external tidal field by exploring the
\textsc{Mocca Survey Database I}, with focus on the presence and evolution
of a stellar-mass black hole subsystem. We argue that the presence of a
black hole subsystem can lead to the dissolution of tidally filling star
clusters and this can be regarded as a third type of cluster dissolution
mechanism (in addition to well-known mechanisms connected with strong mass
loss due to stellar evolution and mass loss connected with the relaxation
process). This third process is characterized by abrupt cluster dissolution
connected with the loss of dynamical equilibrium. The abrupt dissolution is
powered by strong energy generation from a stellar-mass black
hole subsystem accompanied by tidal stripping.
Additionally, we argue that such a mechanism should also work for even
tidally under-filling clusters with top-heavy initial mass function.
Observationally, star clusters which undergo dissolution powered by the
third mechanism would look as a 'dark cluster' i.e. composed of stellar
mass black holes surrounded by an expanding halo of luminous stars
\citep{BanerjeeKroupa2011}, and they should be different from 'dark
clusters' harbouring intermediate mass black holes as discussed by
\citet{Askar2017a}. An additional observational consequence of an
operation of the third dissolution mechanism should be a larger than
expected abundance of free floating black holes in the Galactic halo. 
\end{abstract}




\section{Introduction}\label{Intro}



Understanding the evolution of star clusters, and particularly the way 
in which they lose mass, has been one of the long standing problems of
stellar dynamics. Already in 30s and 40s of the previous century \citet{Ambartsumian1938} and \citet{Chandrasekhar1942} pointed out the role relaxation plays in star cluster evolution. In the 90s
\citet{ChernoffWeinberg1990} showed that star cluster structure, stellar evolution and tidal field of the parent galaxy have a very strong
influence on cluster evolution and its dissolution time. This was a 
seminal study, which later resulted in many more detailed papers 
discussing the final state of star clusters. It was shown that several
other factors can also influence the lifetime of star clusters, such as:
crossing time-scale \citep{Whitehead2013}, primordial binary population
\citep[e.g.][]{TanikawaFukushige2009}, type of the Galactic orbit
\citep[e.g.][]{BaumgardtMakino2003}, the form of the Galactic potential and
tidal shocking \citep[e.g.][]{GnedinOstriker1997}, and the properties of 
dark remnants retained in star clusters \citep{BanerjeeKroupa2011,Contenta2015}.

In this paper, inspired by the works of
\citet{FukushigeHeggie1995,BanerjeeKroupa2011,Whitehead2013} and
\citet{Contenta2015} where cluster dissolution is discussed from the 
point of view of stellar evolution mass loss and formation of a dark subcluster, we decided to take a closer look at the process of dissolution for tidally filling clusters in which a massive subsystem of stellar-mass black holes (BHS) forms and survives until the death of the cluster. 

There are generally two types of cluster dissolution discussed in the literature. Slow, connected with relaxation-driven mass loss, called 'skiing', and abrupt, considered as dynamical, connected with stellar evolution mass loss, called 'jumping' \citep{Contenta2015}. \citet{FukushigeHeggie1995} attributed the final abrupt cluster dissolution to the loss of dynamical equilibrium within the cluster. In such a situation, a cluster will not undergo core collapse\footnote{In the paper, by the term 'core collapse' we will mean the phase of the cluster evolution, in which the energy needed to support cluster structure is provided not through release of potential energy by contracting central cluster parts, but by energy generation, in the cluster core, by binaries or an intermediate mass black hole. It is worth to note that observers use the term 'core collapse' to mean a particular spatial cluster structure that has a central power-law surface brightness profile and cannot be well fitted by a King profile, which is characteristic for 'pre-collapse' clusters. For clusters supported by binaries formed in dynamical interactions or by an intermediate mass black hole, both definitions are equivalent. Only in the case of primordial binaries and BHS they have different meaning.}. Also as it was pointed out by \citet{BanerjeeKroupa2011}, in the case of a strong tidal field, the final stages of cluster evolution can be connected with fast removal of
luminous stars and formation of a so called 'dark cluster' consisting of 
BHs and neutron stars (NSs). They attributed the cluster dissolution to the loss of virial equilibrium for luminous stars and rapid dynamical evolution of a bound very low $N$ cluster composed of evolved stars. 

In this paper we introduce a third mechanism for cluster dissolution. We will investigate processes responsible for the 'jumping' cluster dissolution connected with the formation of a massive BHS and strong energy generation in dynamical interactions between BHs and BH-BH binaries (BBHs). We will examine what is the influence of the cluster initial properties on the cluster lifetime and the way in which it is dissolving.

The structure of the paper is as follows. In Section 2 we describe
the used \textsc{Mocca} and \textsc{N-body}  models. In Section 3 we present our results with some discussion on possible mechanisms leading to final cluster dissolution. The final Section summarizes our conclusions
and tries to relate early cluster dissolution to the possible population
of BHs in the Galactic halo.

\section{Model}\label{Model}

In this paper, we use the results from the \textsc{Mocca-Survey
Database I} \citep{Askar2017b} that contains about 2000
models of globular clusters (GC) with different initial masses and 
structural and orbital parameters. These models were simulated 
with the \textsc{Mocca} code \citep{HG2013}, which treats the relaxation
process using the method described by \citet{Henon1971}, 
that was significantly improved by \citet{Stod1982,Stod1986}, 
and more recently by Giersz and his collaborators 
\citep[and reference therein]{Giersz2008,Giersz2013,Giersz2015,HG2013}. 
\textsc{Mocca} code consists of the following ingredients: \textsc{Sse} 
and \textsc{Bse} codes \citep{Hurley2000,Hurley2002} for treating binary
and stellar evolution, while strong binary-single and binary-binary 
interactions are handled by the \textsc{Fwebody} code \citep{Fregeau2004}, 
stars are escaping from the cluster according to the description 
given by \citet{Fukushige2000}. Escape is no longer instantaneous, 
but takes place after some delay. The parameters of the cluster 
models in the \textsc{Mocca-Survey Database I} are described in Table 1 
in \citet{Askar2017b}. In short, for half of the simulated models, 
supernovae (SNe) natal kick velocities for NSs and 
BHs are assigned according to a Maxwellian distribution, 
with velocity dispersion of 265 $km s^{-1}$ \citep{Hobbs2005}. In the 
remaining cases, BH natal kicks were modified according to the 
mass fallback procedure described by \citet{Belczynski2002}. Metallicities of the models were selected as follows: Z = $0.0002, 0.001, 0.005, 0.006$ and $0.02$. All \textsc{Mocca} models were characterized by a 
\citet{Kroupa2001} initial mass function, with a minimum and maximum
initial stellar mass of $0.08 M_{\odot}$ and $100 M_{\odot}$, 
respectively. The GC models were described by \citet{King1966} 
models with central concentration parameter values $W_0 = 3, 6$ and 
$9$. They had tidal radii ($R_t$) equal to: $30, 60$ or $120$ pc, 
while the ratios between $R_t$ and half-mass radius ($R_h$) 
were $50, 25$ or the model was tidally filling.
The primordial binary fractions were chosen to be: $5\%, 10\%, 30\%$ 
and $95\%$. Models characterized by an initial binary fraction
equal to or lower than $30\%$ had their initial binary eccentricities selected according to a thermal distribution \citep{Jeans1919}, the semi-major axes according to a flat logarithmic distribution, and the mass ratio according to a flat distribution. For models containing a larger binary fraction, the initial binary properties were instead selected according
to the distribution described by \citet{Kroupa1995,Kroupa2011}.
The models consist of $40000$, $100000$, $400000$, $700000$ and $1200000$
objects (stars and binaries). The GCs were assumed to move on a 
circular orbit at Galactocentric distances between $1$ and $50$ kpc. 
The Galactic potential was modelled in the simple point-mass 
approximation, taking as central mass the value of the Galaxy mass
enclosed within the GCs orbital radius. The GC rotation velocity was set to $220$ $km s^{-1}$ for the whole range of galactocentric distances. As it was pointed out by \citet{Askar2017b}, the initial conditions assumed to create the \textsc{Mocca-Survey Database I} were not specifically selected to reproduce the Galactic GC population. Nevertheless, their 
observational parameters calculated at the present-day exhibit a 
remarkably good agreement with Milky Way GCs (see Fig.1 in \citet{Askar2017b}). 

For the purpose of this paper  we mainly selected tidally filling models and models with $95\%$ binary fraction. The choice of such a high value of binary fraction may seem incomprehensible to the reader, especially when the observed GC binary fractions are recalled \citep[of the order of $10\%$]{Milone2012}. However, we would like to point out, that because of a very wide primordial binary period distribution \citep{Kroupa1995,Belloni2017b} most of the
binaries will be quickly destroyed in dynamical interactions, and final binary fraction (at the Hubble time) will be similar to the observed ones. Additionaly, as it was shown in many papers \citep[e.g.][]{Leigh2015,Belloni2017b,Belloni2018,Belloni2019} models with very high binary fractions better describe in GCs: the observed binary fractions and their spatial distributions, the observed binary properties in the field, the dependence of observed binarity as a function of binary mass, the number and spatial distribution of CVs in star clusters. Therefore, we decided to illustrate the discussion of physical mechanisms leading to abrupt cluster dissolution mostly with models with initially very high binary fraction. As it will be clearly shown in Section~\ref{parameter}, also models with an initially small binary fraction are abruptly disrupted.

To check if the \textsc{Mocca} models properly recover the tidally 
filling cluster dissolution, we run an additional \textsc{N-body}  model. The initial conditions for this model were generated from the \textsc{Mocca} snapshot (stars and center of binary mass: radial position, radial and tangential velocities, mass, and for binaries additionally: semi-major axis, eccentricity and second component mass) 
for the initial model of a cluster. The properties of the tidally 
filling model are: $N=100000, W_0=6, R_t=60$ pc, $R_h=8.8$ pc, binary fraction equal to $0.1$ and mass fallback for BH SN kicks. The model was run with the \textsc{Nbody6++gpu} code \citep{Wang2015}. This is a hybrid parallelized version based on the direct $N$-body code \textsc{Nbody6} \citep{AarsethBOOK} designed for the realistic modelling of star clusters. The single and binary stellar evolution packages \textsc{Sse}/\textsc{Bse} \citep{Hurley2000,Hurley2002} are used\footnote{The \textsc{Nbody6++gpu} code version is based on the Master version (Commits on Sept 28, 2018) on GitHub: https:/1/github.com/nbodyx/Nbody6ppGPU/commits/master. The \textsc{Sse}/\textsc{Bse} packages are the original version without recent updates used in \textsc{Nbody6}}. In this \textsc{N-body}  code version, the SNe kicks are treated exactly as in the \textsc{Mocca} code.

It is important to point out that the properties of compact binaries outputted by \textsc{Mocca} and \textsc{Nbody6++gpu} strongly depend on 
the treatment of the common envelope phase in binaries. The version we 
used for our study assumes $\alpha=3$, $\lambda=0.5$ \citep{Hurley2002}, which are default values used in the \textsc{Bse} code. In most N-body  simulations $\alpha=3$, $\lambda=0.0$ are used by default. As a result, a lot of binaries survive to the point where both components collapse into BHs, and a significant number of mergers are primordial, without any strong dynamical interactions \citep{Belloni2017}. \\

\section{Results}\label{Results}

Many theoretical investigations and star cluster simulations have pointed out that for clusters embedded in a galactic tidal field the cluster dissolution depends on the initial cluster relaxation time and the concentration and shape of the cluster initial mass function (IMF). Already, \citet{ChernoffWeinberg1990} showed that a cluster with small concentration (King model parameter $W_0$ equal to 1 or 3) will very quickly dissolve (before any substantial dynamical evolution). This cluster dissolution is powered by strong stellar evolution mass loss coupled with strong cluster expansion and overflow of the $R_t$. Dissolution is characterized by the abrupt loss of equilibrium and very fast dissolution - 'jumping' shape of the evolution of the total cluster mass. Models which are more concentrated
are controlled by the relaxation process, not by stellar evolution mass loss. They evolve much slower and can enter the core collapse phase, provided that the relaxation was fast enough. They are characterized by the so called 'skiing' shape of the total mass evolution. There is a clear interplay between the time-scale of the relaxation process and the time-scale of stellar evolution mass loss, which act differently depending on the cluster concentration, strength of the tidal field and IMF shape.

In this Section, we will discuss the third mechanism leading to the abrupt cluster dissolution, namely formation and strong energy generation by BHS in tidally limiting star clusters. For such models clusters undergo core collapse connected with strong mass segregation and BHS formation, and later, usually after few Gyr of balanced evolution, suddenly undergo abrupt dissolution.

\subsection{Cluster dissolution - global cluster parameters}\label{parameter}

\begin{figure}
\centering
\includegraphics[width=1.0\columnwidth]{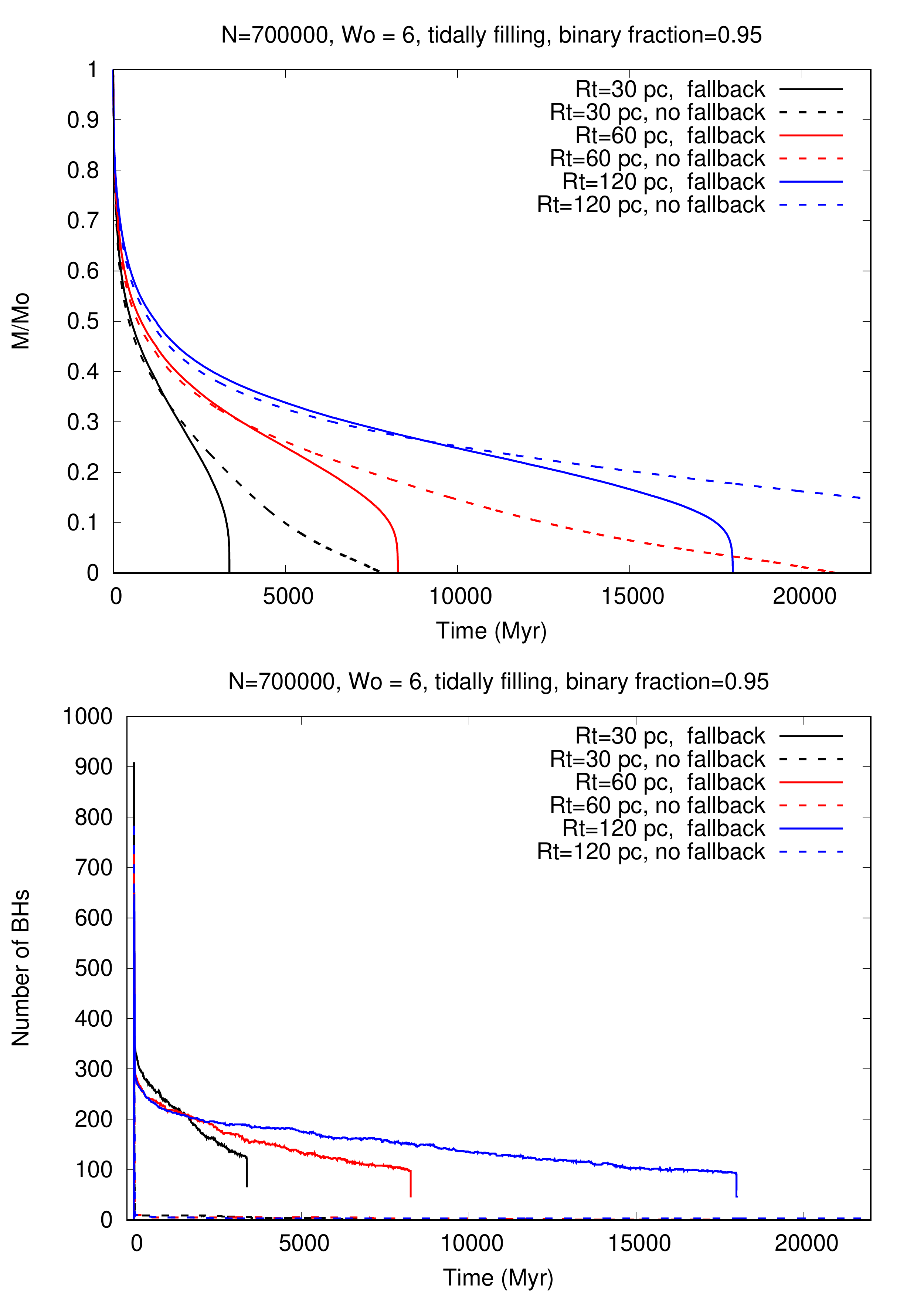}
\caption{Top panel - evolution of the fraction of cluster bound mass as a function of time for tidally filling \textsc{Mocca} cluster models with 700000 objects (stars and binaries), $W_0=6$ and binary fraction equal to $0.95$ for different $R_t$ and SNe natal kicks mass fallback set to ON or OFF. Black line - model with $R_t=30$ pc and mass fallback ON. Black dashed line - model with $R_t=30$ pc and mass fallback OFF. Red line - model with $R_t=60$ pc and mass fallback ON. Red dashed line - model with $R_t=60$ pc and mass fallback OFF. Blue line - model with $R_t=120$ pc and mass fallback ON. Blue dashed line - model with $R_t=120$ pc and mass fallback OFF. Bottom panel - evolution of the number of BHs as a function of time. Line colors and types are as for the top figure.}\label{M-T-all}
\end{figure}

Figure \ref{M-T-all} (top panel) shows the evolution of the fraction of cluster bound mass (normalized to the initial value) for tidally filling models with initially N=$700000$ objects, $95\%$ initial binary fraction, King model concentration $W_0=6$ for different tidal radii and SNe natal kicks with mass fallback ON or OFF. It is clear that models with mass fallback ON evolve much faster than models with mass fallback OFF, and are characterized with abrupt disruption, which in turn suggests loss of the virial equilibrium \citep{FukushigeHeggie1995}. This behaviour is different from the case described by \citet{Contenta2015} where models with NS natal velocity kicks being zero (similar to the \textsc{Mocca} models with fallback ON - reduced natal kicks, but not set to zero) showed clear core collapse and dissolution controlled by the relaxation process, while models with large NS natal velocity kicks (similar to the \textsc{Mocca} models with fallback OFF) showed abrupt dissolution features controlled by strong mass loss due to stellar evolution. This different behaviour is connected with the maximum mass of the main sequence stars for IMF chosen by \citet{Contenta2015}. Limiting the maximum mass to only 15 $M_{\odot}$ prevents formation of BHs which can segregate and form a BHS which takes over the control of global cluster evolution, as it can be seen in Section \ref{BHS}. Figure \ref{M-T-all} (bottom panel) shows the evolution of the number of BHs bound to the system. Models with SNe natal kicks with mass fallback ON show very slow evolution of the BH number. This suggests that BHs close to the final stages of cluster evolution could play a crucial role in the determination of the cluster structure and evolution, as it can be seen in Section \ref{BHS}.

Cluster mass evolution shown in Figure \ref{M-T-all} is very similar until the mass fraction is about 0.7 for all models irrespective of the $R_t$, or mass fallback option. This strongly suggests that the initial cluster evolution is governed by mass loss connected with the stellar/binary evolution of massive stars. Then the relaxation process takes over and the cluster evolution starts to depend on $R_t$. The larger the $R_t$, the longer the relaxation time and the slower the cluster evolution. From the fraction of cluster bound mass about 0.3 the evolution of cluster with mass fallback ON and OFF starts to differ. Models with fallback ON evolve faster. BHS takes over and governs the cluster evolution and finally leads to the loss of equilibrium within the cluster. 

\begin{figure}
\centering
\includegraphics[width=1.1\columnwidth]{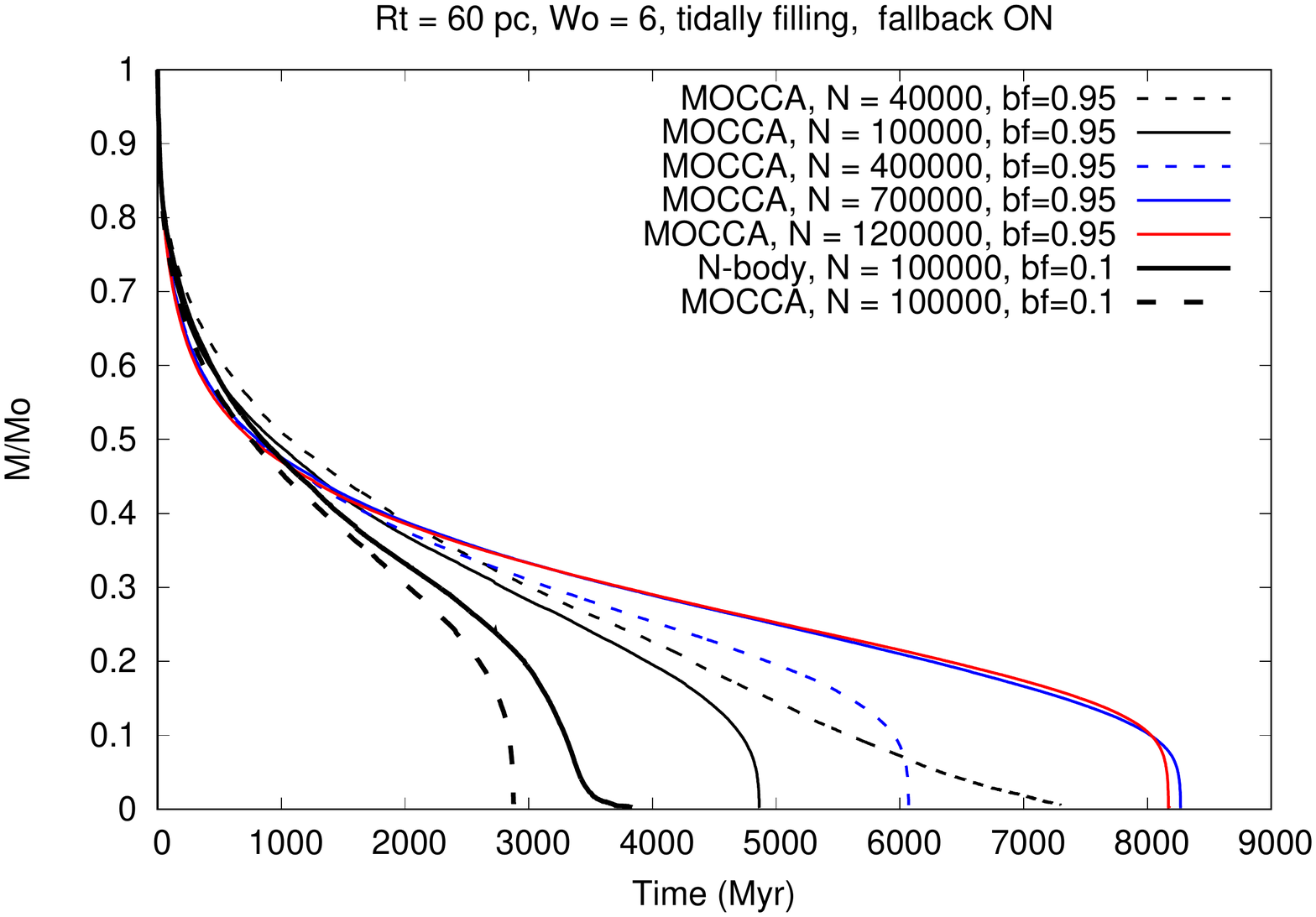}
\caption{Evolution of the fraction of cluster bound mass as a function of time for the \textsc{Mocca} tidally filling models with $R_t=60$ pc, $W_0=6$, binary fraction $0.95$, mass fallback ON and different number of objects N. Black dashed line - $40000$. Black line - $100000$.  Blue dashed line - $400000$. Blue line - $700000$. Red line - $1200000$.
Additionally, models with $100000$ objects and binary fraction $0.1$ are depicted. Thick black dashed line - \textsc{Mocca} model. Thick black line - \textsc{N-body} model.}\label{M-T-N}
\end{figure}

To check if the abrupt cluster dissolution shown in the \textsc{Mocca} simulations is real, and not connected with the physical assumptions underlying the Monte Carlo method, we decided to run a direct \textsc{N-body} model and check if it will show the same behaviour as the \textsc{Mocca} model. We have chosen a small N model with a small binary fraction, namely $N=100000$ and $10\%$ binary fraction. Such models are relatively easy for \textsc{N-body} simulations. The model was run with the \textsc{Nbody6++gpu} code \citep{Wang2015}. We used the initial \textsc{Mocca} snapshot (positions, velocities and masses) to setup the initial conditions for the run. As it can be seen from Figure \ref{M-T-N}, for $N=100000$ models with thick black lines, the \textsc{N-body} model shows the same behaviour as the \textsc{Mocca} model. The small differences observed from the very beginning are connected with slightly different prescriptions in the \textsc{Bse} code for the BH masses formed in SNe and the amount of accreted mass on BHs during collisions/mergers with other stars. The final cluster dissolution in the  \textsc{N-body} model is slightly less abrupt than in the \textsc{Mocca} model, but we should remember that the Monte Carlo method is less accurate for processes which act in time-scales proportional to the dynamical time-scale. This is the case for the final cluster dissolution. Concluding, we can safely assume that the abrupt dissolution of clusters observed in the \textsc{Mocca} simulations are real features of cluster evolution. It is important to stress here that abrupt dissolution means a final loss of about $15\%$ of the initial cluster mass in about a few hundred Myr, compared to about dozens of Gyr of slow mass loss powered by the relaxation process.

\begin{figure}
\centering
\includegraphics[width=1.1\columnwidth]{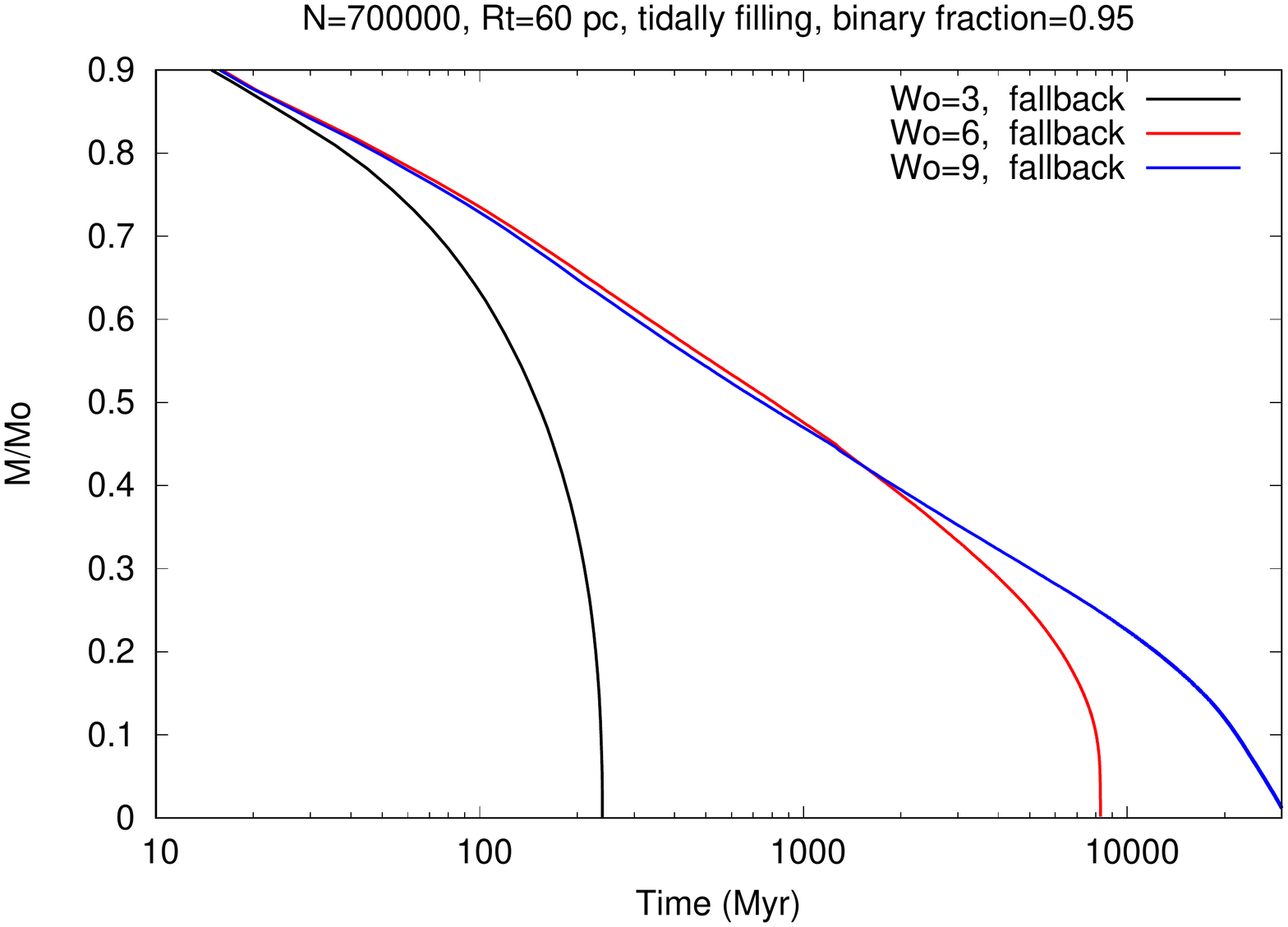}
\caption{Evolution of the fraction of cluster bound mass as a function of time for the \textsc{Mocca} tidally filling models with 700000 objects (stars and binaries), $R_t=60$ pc, binary fraction $=0.95$, mass fallback ON and different $W_0$. Black line - $W_0 = 3$. Red line - $W_0 = 6$.  Blue line - $W_0 = 9$.}\label{M-T-Wo}
\end{figure}

\begin{figure}
\centering
\includegraphics[width=1.1\columnwidth]{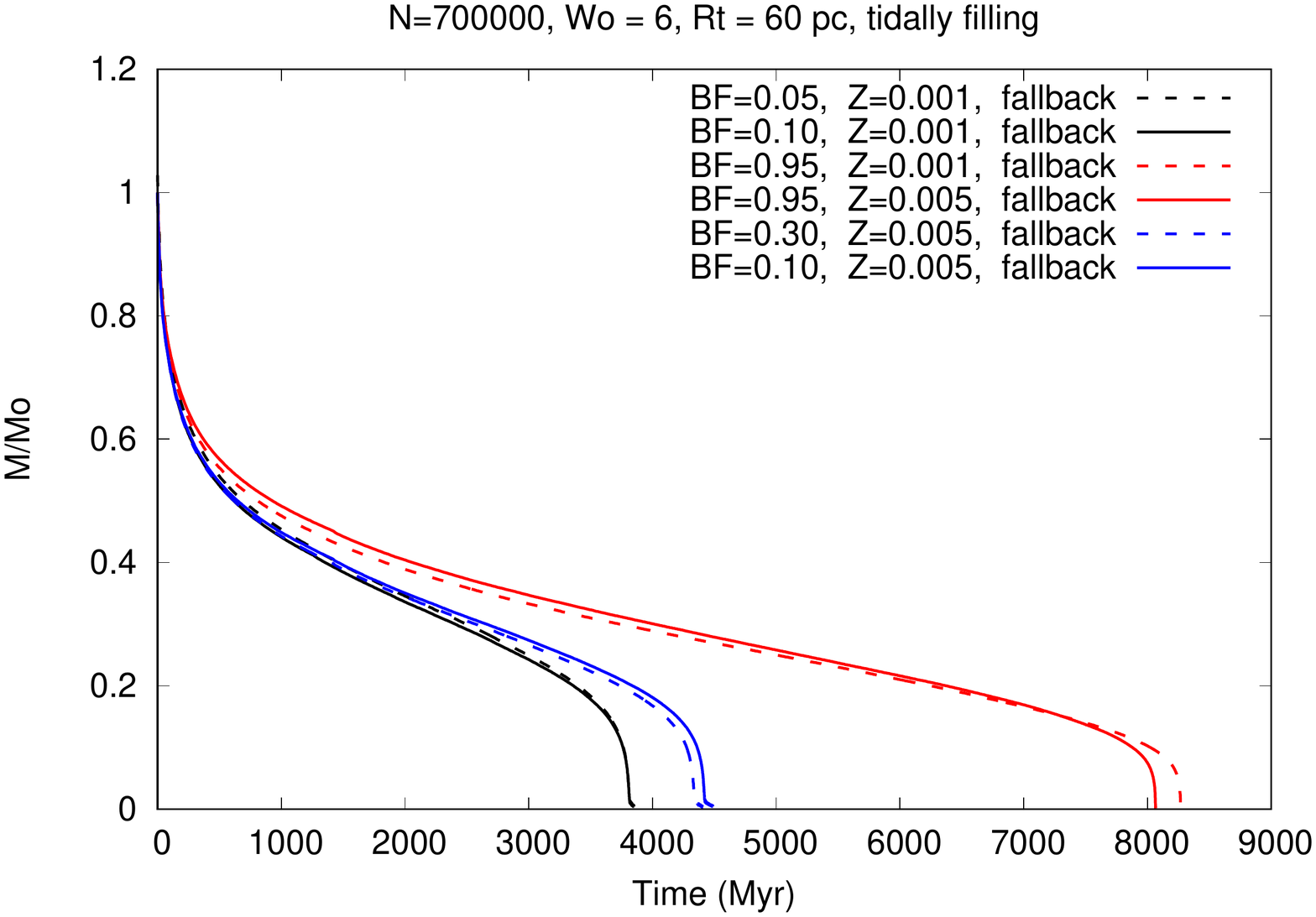}
\caption{Evolution of the fraction of cluster bound mass as a function of time for the \textsc{Mocca} tidally filling models with 700000 objects (stars 
and binaries), $R_t=60$ pc, $W_0=6$, mass fallback ON and different 
binary fractions (BF) and metallicities (Z). Black dashed line - BF = $0.05$ and Z = $0.001$. Black line - BF - $0.1$ and Z = $0.001$. Red dashed line - BF = $0.95$ and Z = $0.001$. Red line - BF = $0.95$ and 
Z = $0.005$. Blue dashed line - BF = $0.3$ and Z = $0.005$. Blue line - 
BF = $0.1$ and Z = $0.005$.}\label{M-T-bfz}
\end{figure}

\begin{figure}
\centering
\includegraphics[width=1.1\columnwidth]{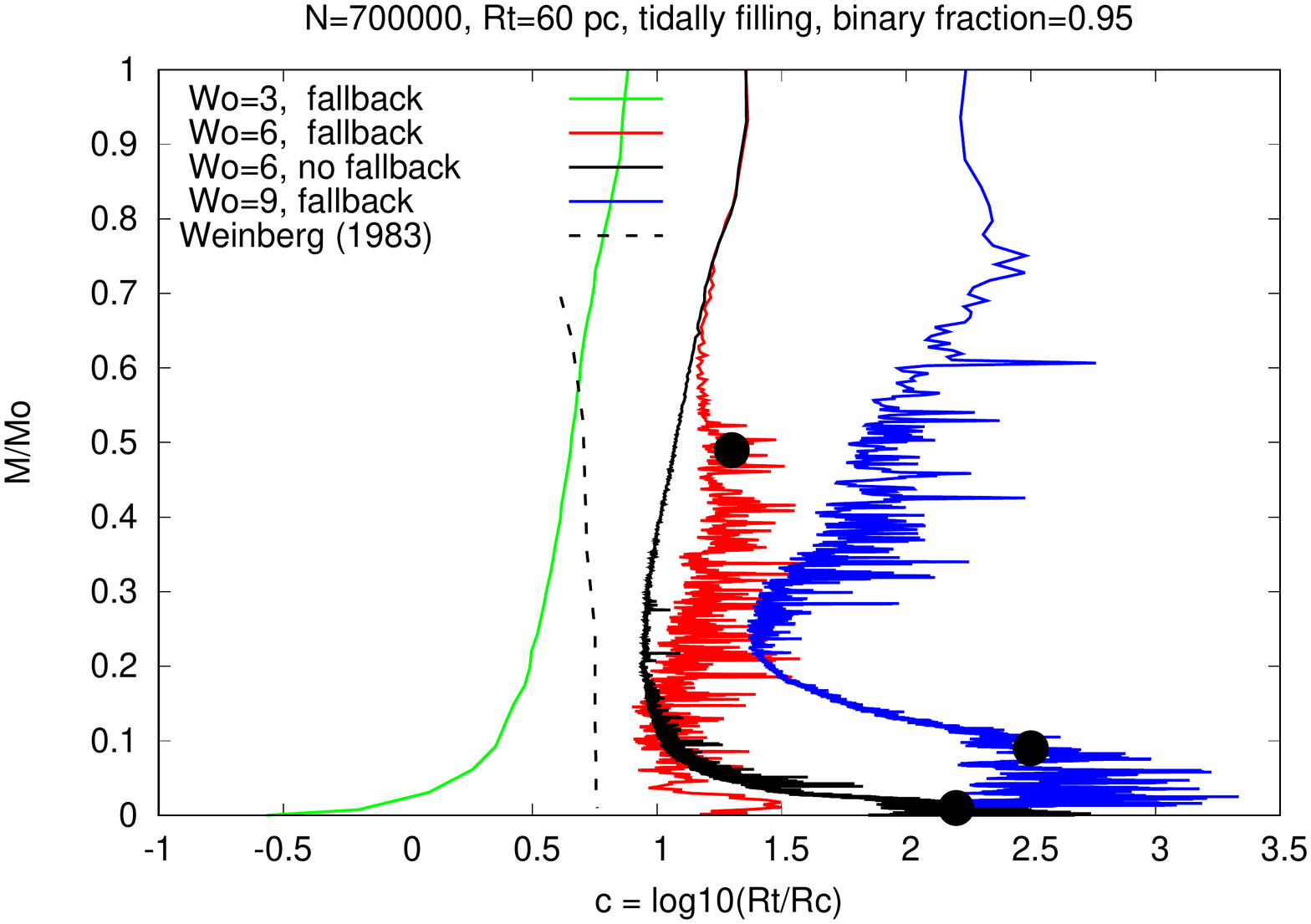}
\caption{The fraction of cluster bound mass as a function of the concentration parameter for the \textsc{Mocca} tidally filling models with 700000 objects (stars 
and binaries), $R_t=60$ pc, binary fraction = $0.95$, tidally filling models with mass fallback ON for different $W_0$. Green line - $W_0=3$. Red line - $W_0=6$. Black line - $W_0=6$ and mass fallback OFF. Blue line - $W_0=9$. The black dashed vertical line shows the "Weinberg cliff" (see Figure 3 in \citet{Weinberg1993}). The black dots mark roughly the moment of the core collapse.}\label{W-c-M}
\end{figure}

In the literature, in order to describe the local and global cluster evolution rates, there are generally used definitions of relaxation times given in Spitzer's book \citet[][Eqs. 2-62 and 2-63 respectively]{Spitzer1987}. The definition of local relaxation time is as follows: 
\begin{equation}\label{eq:1}
T_r = \frac{0.065~v_m^3}{G~\rho~m~ln\Lambda},
\end{equation}
where G is the gravitational constant, $v_m$ is the velocity dispersion, $\rho$ is the mass density, $m$ is the average mass and $ln\Lambda$ is the Coulomb logarithm. The half-mass relaxation time if given by:
\begin{equation}\label{eq:2}
T_{rh} = \frac{0.138~M~R_h^{3/2}}{G^{1/2}~m~ln\Lambda},
\end{equation}
where $M$ is the cluster mass. As it can be easily seen from Eqs.~\ref{eq:1} and \ref{eq:2}, the larger the cluster mass or the half-mass radius, the smaller the density or the average mass, the longer the relaxation time. Those dependences will be explored during discussions of the Figures presented in this Section.
 
In Figures \ref{M-T-N}, \ref{M-T-Wo} and \ref{M-T-bfz}, the evolution of the fraction of cluster bound mass is shown as a function of time for models with different: number of objects, King model concentrations, binary fractions and metallicities, respectively. As it was already shown during discussion of Figure \ref{M-T-all} most of the cluster evolution is governed by the relaxation process. Therefore, it is no surprise that models which are characterized by shorter half-mass relaxation time evolve faster and earlier enter the abrupt dissolution phase. Shorter relaxation time means: smaller number of objects (see Figure \ref{M-T-N}), smaller King concentration parameter leading to faster mass loss and stronger tidal striping (see Figure \ref{M-T-Wo}) and smaller binary fraction, (see Figure \ref{M-T-bfz}). Models with very large binary fraction, $0.95$, are initially characterized by very wide period distribution \citep{Kroupa1995}, so a large number of binaries are quickly disrupted in dynamical interactions and the number of objects is sharply increased leading to longer relaxation times. The dependence of the cluster disruption time on the cluster metallicity (see Figure \ref{M-T-bfz}) is negligible, which is not surprising. As we know from stellar evolution models, stars with larger metallicities have stronger winds and lose more mass before BH formation \citep{Vink2001}. So, BHs have slightly smaller masses and clusters should have slightly larger half-mass radii, because of enhanced stellar evolution mass loss. Larger half-mass radius means slower evolution (see Eq.~\ref{eq:2}) and smaller BH masses mean denser central parts of systems and faster removal of BHs from the system in dynamical interactions. Both effects leads to opposite behaviours, and they nearly cancel each other out. So, the abrupt cluster dissolution time does not strongly depend on metallicity.  

There are only three exceptions from the picture discussed above. In Figure \ref{M-T-N} we see that the model with the smallest number of objects, $N=40000$, does not show abrupt dissolution. This model is characterized with the shortest initial relaxation time (among models considered in this paper), about 1.6 Gyr, and the smallest number of retained BHs, about 30. In such a situation, according to \citet{BreenHeggie2013}, BHs are quickly kicked out from the system and BHS cannot survive for a long time. So, the cluster evolution is not governed by strong energy generation by BHs in BHS. In Figure \ref{M-T-Wo} we can see that the model with $W_0=3$ dissolves extremely fast and the model with $W_0=9$ does not show the abrupt dissolution feature. The very fast dissolution of tidally filling models with low King model concentration was already extensively discussed in the literature \citep[e.g.][]{Weinberg1993,FukushigeHeggie1995,Whitehead2013,Contenta2015}. The dissolution of such clusters is controlled by very strong initial mass loss powered by stellar/binary evolution. Relaxation process is not important at all. The situation is much different for the model with $W_0=9$. As it can be seen from Figure \ref{W-c-M}, the cluster enters the core collapse phase, so it has to be controlled by the relaxation process. The difference between the model with $W_0=6$ and the model with $W_0=9$ is connected with the rate of BHS evolution. Model $W_0=9$ is initially much denser, so its half-mass radius and half-mass relaxation time are shorter than for the $W_0=6$ model. Nevertheless, for both models the BH mass segregation ends nearly at the same time, about 4 Gyr. At that time, the models contain about 50 BHs and 160 BHs for $W_0=9$ and $W_0=6$, respectively. According to \citet{BreenHeggie2013} the evolution of BHs is controlled by the energy flow through the cluster half-mass radius, which is proportional to $E_b/T_{rh} \approx GM/R_h/T_{rh}$, where $E_b$ is the cluster binding energy. Models with $W_0=9$ have smaller half-mass radius and half-mass relaxation time than models with $W_0=6$, so the energy demand is larger for this model and leads to a much faster burning out of BHs - larger number of dynamical interactions leading to BH removal from the system. After the time of BH mass segregation (about 4 Gyr) the model $W_0=6$ contains enough BHs to form a BHS and enters the phase of balanced evolution \citep{BreenHeggie2013}. Contrarily, the model with $W_0=9$ has too small a number of BHs and continues to remove BHs quickly to support the needed energy flow. Finally, it enters the phase when other energy sources connected with ordinary binaries take over and the cluster enters the core collapse and then core bounce phases. It is important to note that the tidal field plays an important role in the above picture. The less concentrated model, has larger $R_h$ and loses more mass, so it is easier for the BHS to provide the needed energy to support the cluster structure. This phase of evolution ends when the cluster suddenly loses its virial equilibrium.

There is another feature suggesting very strong influence of the BHS on the abrupt cluster dissolution. As it is shown in Figure \ref{M-T-N}, models with initially $N=700000$ and $N=1200000$ objects dissolve nearly at the same time, whereas other models with smaller $N$ follow more or less the dependence on the relaxation time - the shorter the relaxation time, the faster the cluster evolution. So, model $N=1200000$ speeds up its evolution. The faster evolution is connected with a more massive BHS and larger energy generation, which leads to stronger mass loss, additionally increased by the tidal stripping. We have to remember that the more massive cluster model has initially the same tidal and half-mass radii as the lower mass model, so the energy flow at $R_h$ is much larger and the BHS has to evolve faster to provide the needed energy. To a smaller extent this is also visible for the smaller N models.

At the end of this section we would like to discuss Figure \ref{W-c-M}, in which the evolution of the fraction of cluster bound mass as a function of the King concentration parameter for models with different $W_0$ and type of mass fallback is depicted. As it was already pointed out by \citet{Weinberg1993}, models on the left of the vertical black line (so called 'Weinberg's cliff') are dissolving on the dynamical time-scale because of the strong mass loss connected with stellar evolution. Models on the right of this line evolve on the relaxation time scale, and usually before dissolution they undergo core collapse. For the \textsc{Mocca} models this is true with the exception of the tidally filling model with $W_0=6$ and with mass fallback ON. For this model, despite the evolution time-scale being proportional to the relaxation time and the relatively early core collapse, it enters an abrupt dissolution phase and very quickly dissolves, still being on the right side of the 'Weinberg's cliff'. As we argued above, this dissolution is connected with the formation and evolution of a BHS in strongly tidally striped clusters. This model is an example of the third dissolution mechanism. 

\subsection{Cluster dissolution - black hole subsystem properties and evolution}\label{BHS}

In the previous Section we showed that the BHS seems to play a crucial role in the dissolution of the tidally filling star clusters. Now, we will try to understand why that is by looking at the BHS evolution and its properties. It is well known from the theoretical paper by \citet{BreenHeggie2013} and simulation results discussed by \citet{ArcaSeddaetal2018}, \citet{Askaretal2018} and \citet{Krameretal2019} that the evolution of the BHSs depends strongly on the cluster half-mass relaxation time and it is connected with the continuous removal of BHs in dynamical interactions - first most massive and then less massive ones. Such a decreasing number of BHs is accompanied with smaller BHS size, larger BHS characteristic density, smaller average BH masses, higher fraction of BHs in binaries and larger average mass of ordinary stars mixed with BHs in the BHS. Using these relations we will try to understand the reason for tidally filling cluster dissolution by BHSs.

\begin{figure}
\centering
\includegraphics[width=1.0\columnwidth]{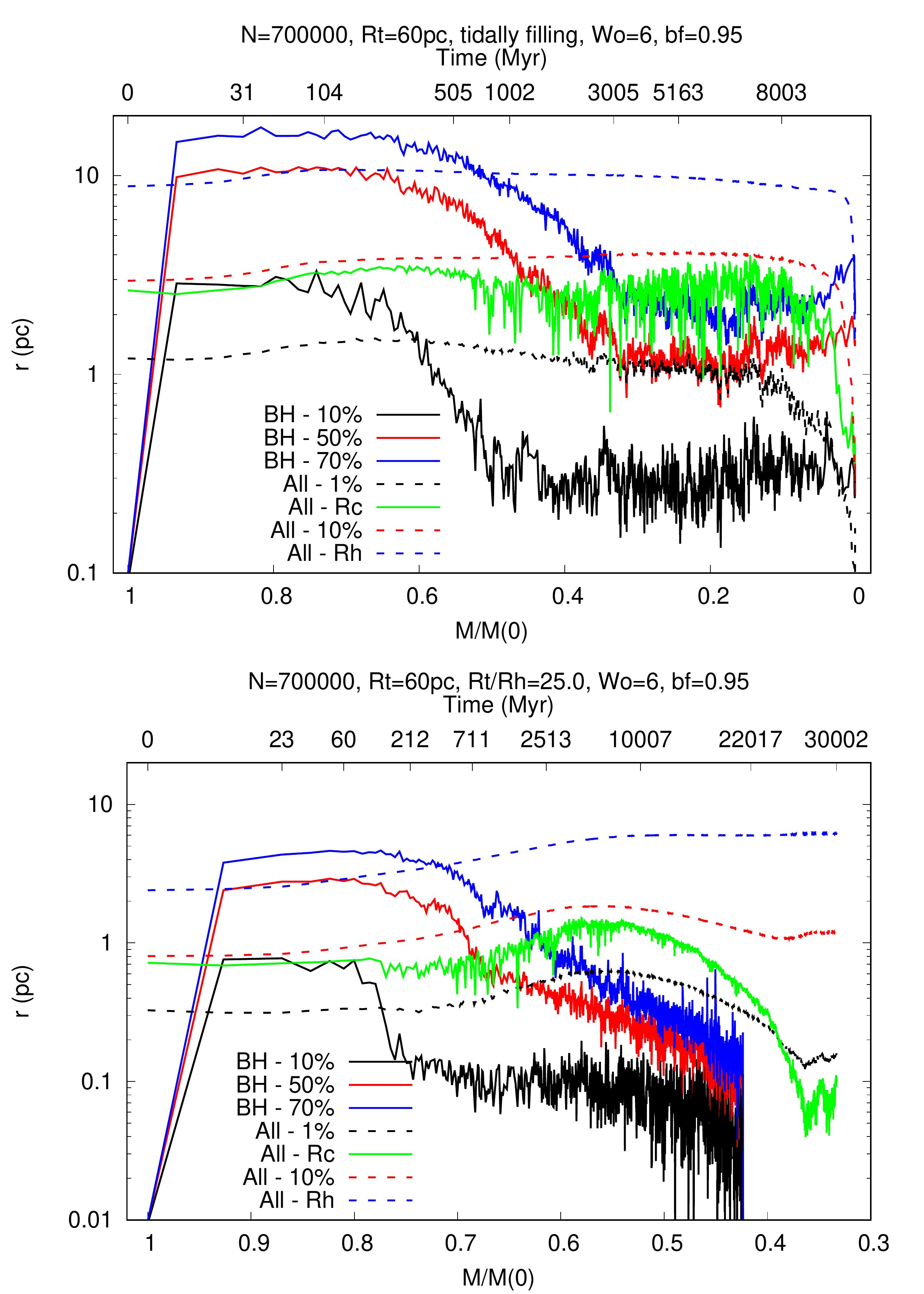}
\caption{Evolution of the cluster Lagrangian radii as a function of 
the fraction of cluster bound mass for the \textsc{Mocca} tidally filling model (top panel) and tidally under-filling model ($R_t/R_h=25$, bottom panel) with  $700000$ objects (stars and binaries), $R_t=60$ pc, $W_0=6$, binary fraction = $0.95$. Black line - BH $10\%$ Lagrangian radius. Red line - BH $50\%$ Lagrangian radius. Blue line - BH $70\%$ Lagrangian radius. Black dashed line - overall $1\%$ Lagrangian radius. Green line - overall core radius. Red dashed line - overall $10\%$ Lagrangian radius. Blue dashed line - overall half-mass radius. On the top axes, cluster evolution times are shown to give the reader an information about the cluster evolution rates.}\label{R-M-tf}
\end{figure}

Figures \ref{R-M-tf}  show the evolution of the overall Lagrangian radii and BHS Lagrangian radii for tidally filling and tidally under-filling models. From the inspection of the Figure we can clearly see the initial overall cluster expansion powered by the stellar evolution mass loss. For the tidally under-filling model, expansion is stronger than in the tidally filling model, because for such a model, the cluster can freely expand until it reaches the tidal radius. Then, for both models we can observe the mass segregation of BHs. For the tidally filling model it stops for the innermost $10\%$ Lagrangian radius at about 0.5 for the fraction of cluster bound mass, and for the tidally under-filling model at about 0.75 for the fraction of cluster bound mass. The mass segregation for the other BH Lagrangian radii continues, and it stops at about 0.3, or never stops for tidally filling and tidally under-filling models, respectively.
For the tidally filling model we can clearly see that BH mass segregation is connected with the expansion of the central parts of the cluster (except $1\%$ overall Lagrangian radius, which is dominated by BHs) and decreasing outer overall Lagrangian radii (see the $50\%$ Lagrangian radius), which is connected with strong tidal striping of the cluster. This behaviour, as we will see later, seems to be very important for understanding the final abrupt cluster dissolution. For the tidally under-filling model, the inner overall Lagrangian radii starts to decrease when the BHS contracts to support cluster structure and needs to generate more energy  (decrease of the $1\%$ BHS Lagrangian radius). Half-mass radius is still increasing in accord with central energy generation and the tidal stripping is not too strong.
 
\begin{figure}
\centering
\includegraphics[width=1.1\columnwidth]{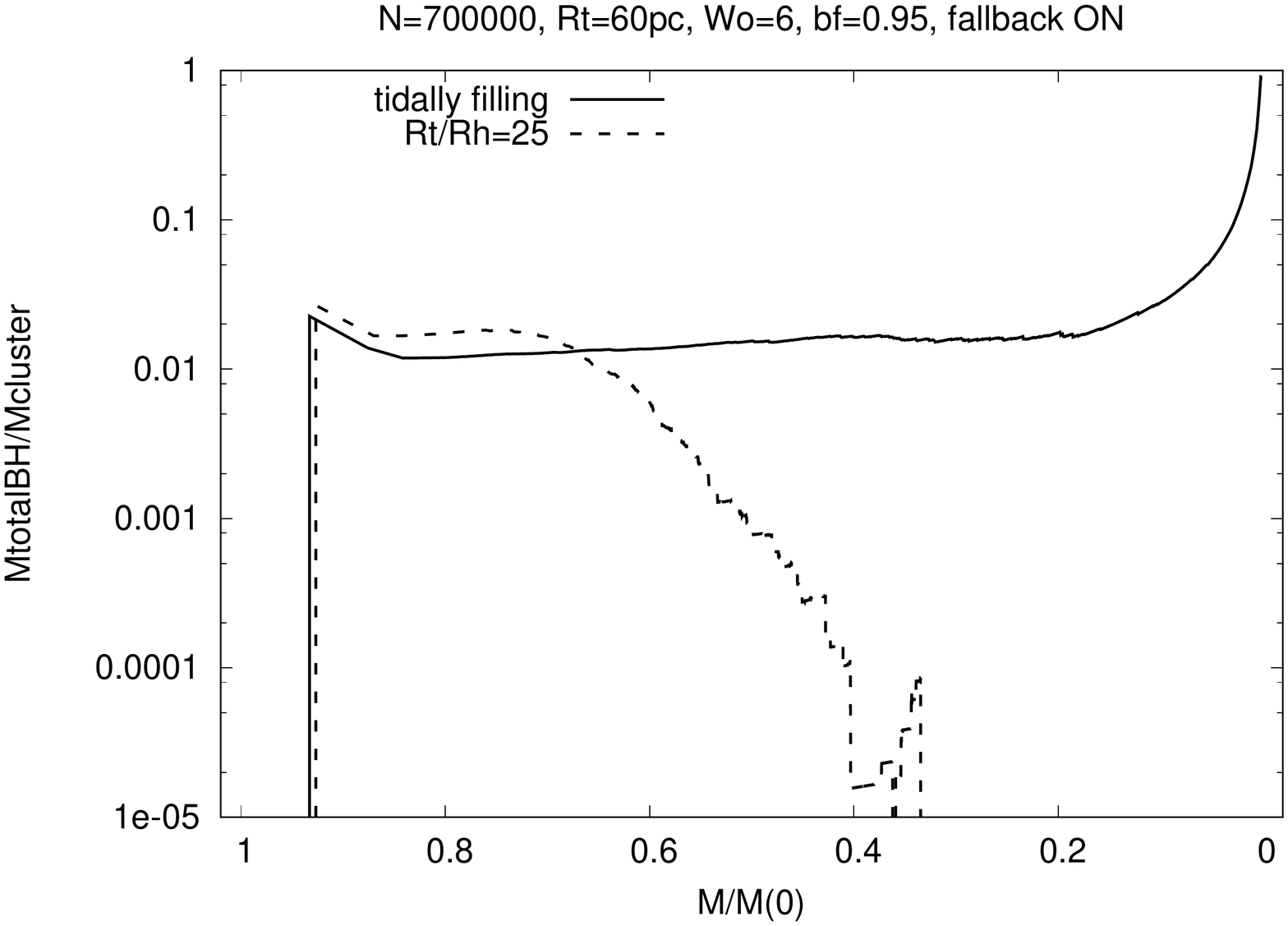}
\caption{Evolution of the ratio between BHS total mass and cluster total mass as a function of the fraction of cluster bound mass  for the \textsc{Mocca} tidally filling and tidally underfiling models with 700000 objects (stars and binaries), $R_t=60$ pc, $W_0=6$, binary fraction $=0.95$ and mass fallback ON. Black line - tidally filling model, Black dashed line - tidaly under-filling model with $R_t/R_h=25$.}\label{MBH-M}
\end{figure}

\begin{figure}
\centering
\includegraphics[width=1.1\columnwidth]{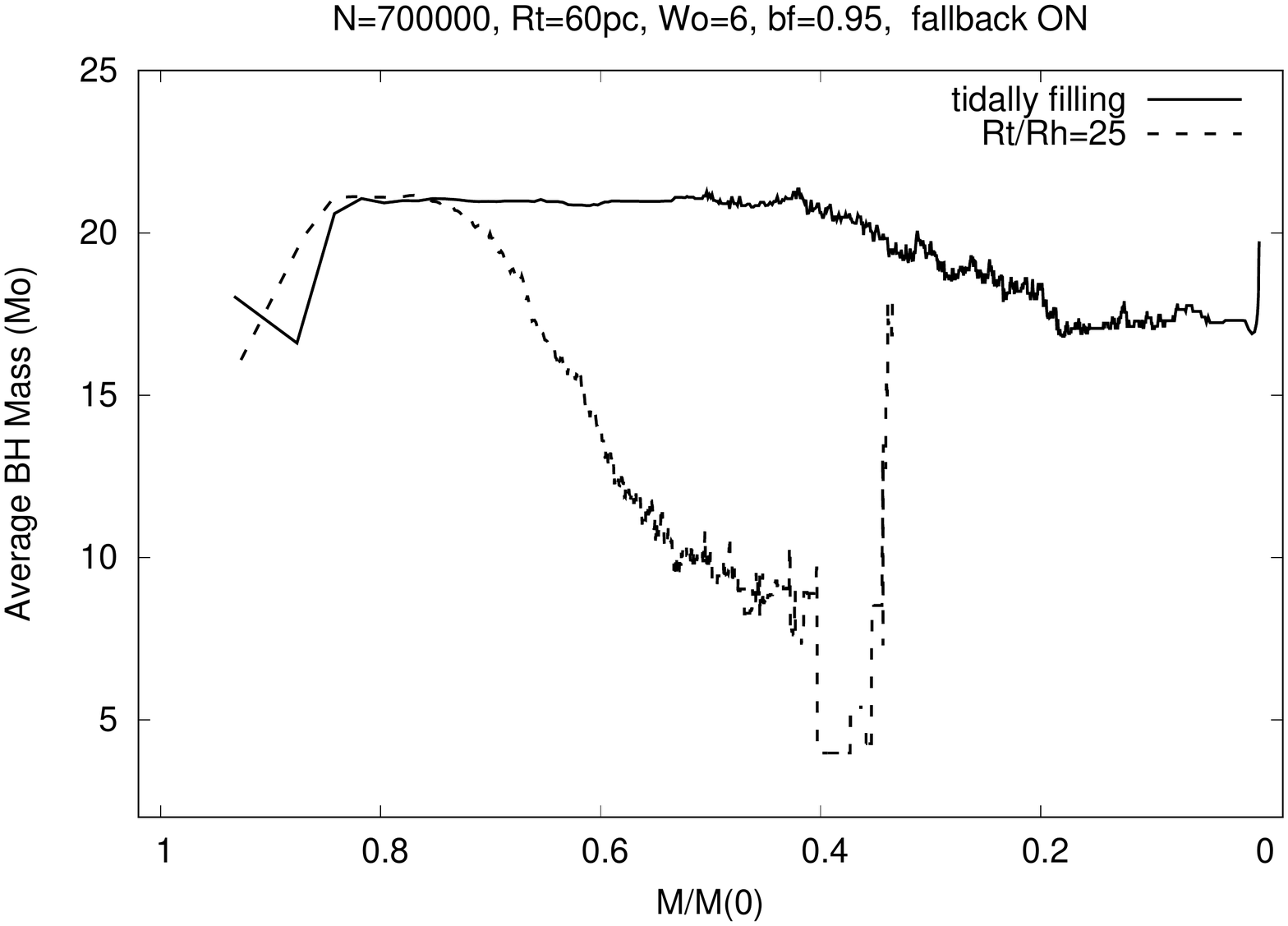}
\caption{Evolution of the BH average mass as a function of the fraction of cluster bound mass  for the \textsc{Mocca} tidally filling and tidally underfiling models with 700000 objects (stars and binaries), $R_t=60$ pc, $W_0=6$, binary fraction $=0.95$ and mass fallback ON. Black line - tidally filling model, Black dashed line - tidaly under-filling model with $R_t/R_h=25$.}\label{ABH-M}
\end{figure}

The described behaviour above can be easily understood on the basis of the theoretical arguments provided by \citet{BreenHeggie2013} and results of simulations discussed by \citet{ArcaSeddaetal2018,Askaretal2018}. The first part of the balanced evolution of the BHS is characterized by a nearly constant number of BHs and a nearly constant ratio between BHS mass and the cluster mass. The duration of this phase depends on the time-scale of BH mass segregation. When the most massive BHs are kicked out in dynamical interactions the mass of the BHS is decreasing together with the average BH mass (see Figures \ref{MBH-M} and \ref{ABH-M}). The duration of this phase of evolution depends on the half-mass relaxation time. The shorter the half-mass relaxation time, the faster the BHS evolution, the faster the decrease of the average BH mass, and the faster the increase of the BHS density (smaller innermost Lagrangian radius). This general picture is slightly modified for the tidally filling model for the fraction of cluster bound mass smaller than 0.4 and larger than 0.2, when the mass ratio between BHS mass and the total cluster mass is nearly constant instead of decreasing. Such departure from the standard theoretical picture \citep{BreenHeggie2013} can be attributed to the strongly decreasing total cluster mass due to tidal striping. The required energy flow through the half-mass radius is constantly decreasing, so the BHS evolves slower and its mass is changing nearly in accord with the cluster mass. There is some decrease of the average BH mass, due to the removal of the most massive BHs in dynamical interactions. At the point when the the fraction of cluster bound mass is equal to about 0.2, the BHS starts to decouple from the rest of the system. The average BH mass is nearly constant, which suggests a very small rate of energy generation by BHS in dynamical interactions. The BHS Lagrangian radii starts to slowly increase and the cluster stars are quickly tidally stripped, which suggests a loss of dynamical equilibrium (see in Figure \ref{MBH-M} the steep increase of the ratio between BHS mass and the cluster mass).

\begin{figure}
\centering
\includegraphics[width=1.0\columnwidth]{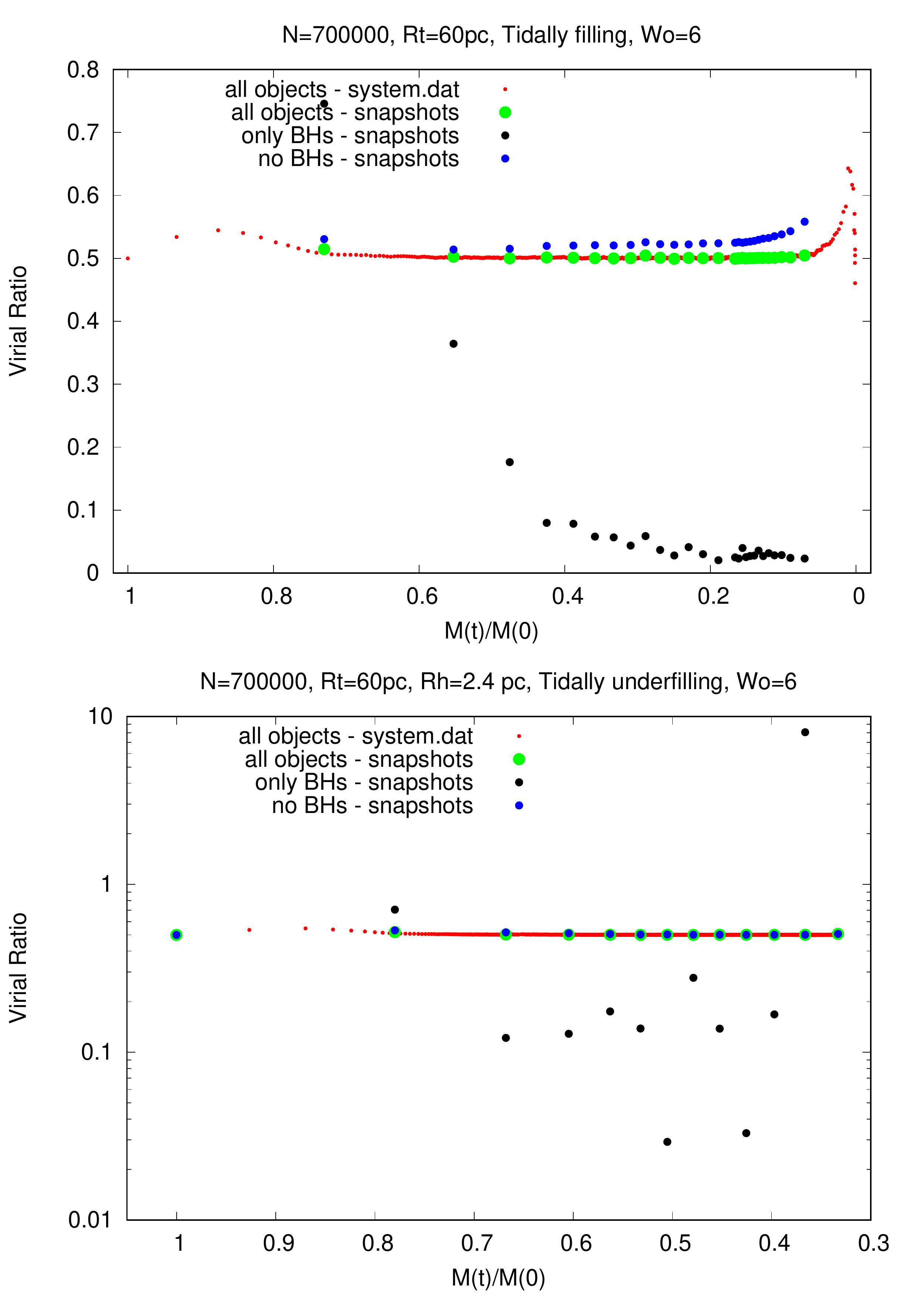}
\caption{Evolution of the virial ratios as a function of the fraction of cluster bound mass for the \textsc{Mocca} tidally filling model (top panel) and tidaly under-filling model ($R_t/R_h=25.0$, bottom panel), with 700000 objects (stars and binaries), $R_t=60$ pc, $W_0=6$, binary fraction $=0.95$, mass fallback ON. Red points - computed from the actual system kinetic and potential energies. Green points - computed from the model snapshots for all objects (only for individual times). Black points - computed from the snapshots only for BHs. Blue points - computed from the snapshots only for objects different than BHs,}\label{Q-M}
\end{figure}

Already \citet{FukushigeHeggie1995,Whitehead2013} and \citet{Contenta2015} showed that the abrupt cluster dissolution powered by strong mass loss due to stellar evolution is connected with the overall loss of cluster dynamical equilibrium. \citet{BanerjeeKroupa2011} in turn showed, for small N clusters, that so the called 'dark clusters' are formed from the BHS, which forces the rest of the system to abrupt dissolution, because of the loss of dynamical equilibrium for the luminous objects (see their Figure 1, top panel). Figure \ref{Q-M} shows the virial ratios computed for the whole system, for BHS and other objects, for the tidally filling model (top panel) and the tidally under-filling model (bottom panel), respectively. In order to compute virial ratios for BHS and other objects from the \textsc{Mocca} models we needed to use the detailed snapshots which are less frequently stored (every 100 Myr) than the standard output (every few Myr). Therefore, there are larger gaps between points for virial ratios computed from the snapshots. From Figure \ref{Q-M} (top panel) we can clearly see that the cluster dissolution for tidally filling model is indeed connected with the decoupling of the BHS from the rest of the cluster and the loss of dynamical equilibrium by other objects. Unfortunately, because of the scarcity of snapshots very close to the cluster dissolution time, we cannot observe the final values of the virial ratios for all cluster components. We also clearly see, that the BHS stops to collapse further at around the fraction of cluster bound mass equal to about 0.2, which coincidences with the slow increase of the BHS Lagrangian radii. The BHS starts to disrupt itself. Its evolution follows the well known behaviour of low-N systems. For the tidally under-filling cluster (bottom panel) we clearly see mass segregation of the BHs and formation of a BHS, but at the fraction of cluster bound mass equal to about 0.65, further decrease of the BHS virial ratio is stopped. It seems that this is connected with the faster removal of the most massive BHs in dynamical interactions than for tidally filling models, and is accompanied by the decrease of BHS Lagrangian radii. The large fluctuations observed for the BHS virial ratios are connected with the decreasing number of BHs. Figure \ref{Q-M} clearly confirms findings by \citet{BanerjeeKroupa2011} and shows that a 'dark cluster' can be formed also for star clusters of the size of globular clusters, not only for open clusters.

\begin{figure}
\centering
\includegraphics[width=1.1\columnwidth]{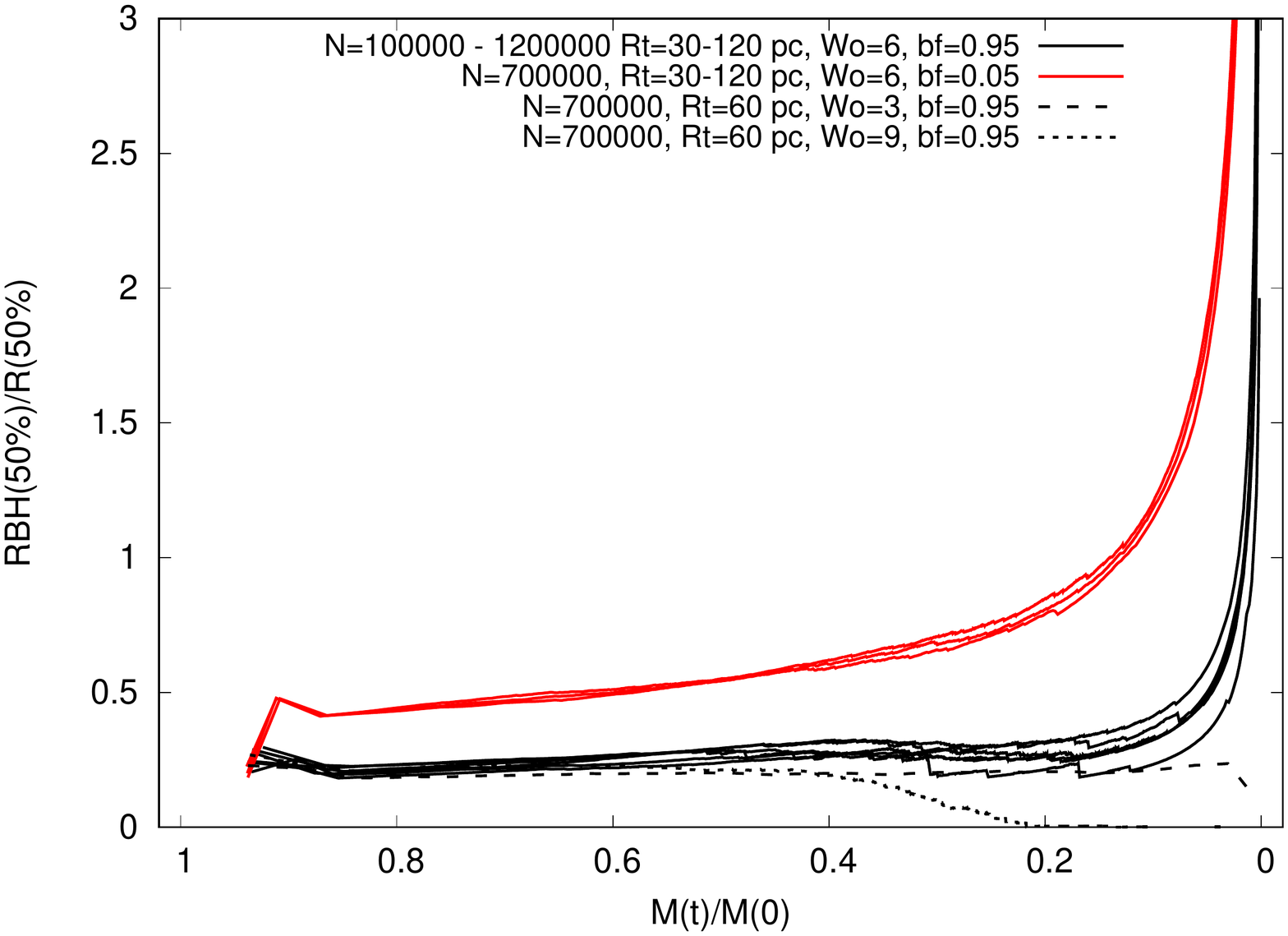}
\caption{Evolution of the ratio between BHS half-mass radius and overall half-mass radius as a function of the fraction of cluster bound mass. The lines are computed using equation 4 in \citet{BreenHeggie2013} for different \textsc{Mocca} models. Black lines - models with number of objects $N$ between $100000$ and $1200000$, $R_t$ between $30$ and $120$ pc, $W_0=6$ and binary fraction $0.95$. Red lines - models with number of objects $N=700000$, $R_t$ between $30$ and $120$ pc, $W_0=6$ and binary fraction $0.1$. Black dashed line - model with number of objects $N=700000$, $R_t=60$ pc, $W_0=3$ and binary fraction $0.95$.
Black doted line - model with number of objects $N=700000$, $R_t=60$ pc, $W_0=9$ and binary fraction $0.95$.}\label{heggie}
\end{figure}

\subsection{Cluster dissolution - Global similarities}\label{global}

\citet{BreenHeggie2013} in their model approximated a star cluster harbouring a BHS by a simple two-component model. They assumed, following the H{\'e}non principle \citep{Henon1975}, that the rate of energy generation by BHS is regulated by the energy demand of the bulk of the system, which is usually thought of as the energy flux at the half-mass radius. Using this approximation, they derived a formula, which relates the half-mass radius of the BHS to the overall cluster half-mass radius. Using equation 4 from \citet{BreenHeggie2013} and the parameters of the BHS and the clusters from the \textsc{Mocca} simulations we plotted on Figure \ref{heggie} the ratio between the half-mass radius of the BHS to the overall cluster half-mass radius for \textsc{Mocca} models, characterized by a different: number of objects, binary fractions, tidal radii and $W_0$. The model parameters are described in the Figure caption. Except models with $W_0=3$ and $W_0=9$, all other models are grouped in two separate families. First, connected with all models having initial binary fraction equal to $95\%$, regardless of the number of objects or $R_t$. Second, connected with models having small binary fractions, also regardless of the number of objects and $R_t$. As it was already discussed before, the model with $W_0=3$ evolves and dissolves very quickly, so the BHs are unable to mass segregate, and therefore the ratio between the BHS half-mass radius and overall half-mas radius is nearly constant. Model with $W_0=9$ is initially very dense and has very short half-mass relaxation time, so the BHS is formed and burned out quickly, therefore the ratio between half-mass radii is decreasing. The observed differences between models with a small and large binary fraction are connected with the fact that small binary fraction models are initially able to retain a larger fraction of BHs on the cluster unit mass. This rather unexpected conclusion (taking into account the fact that models with a very large binary fraction contain nearly twice the number of stars compared to models with a low binary fraction) is connected with the fact of very different initial binary properties in both models. Models with a large binary fraction follow period and mass ratio distributions according to \citet{Kroupa1995}. For such models in the \textsc{Mocca}-Survey Database, the mass ratio is practically one for MS stars which can form BHs, and binary periods can be very large. 
For such binary parameters most of the wide binaries will be disrupted, even for the massive MS stars, by two successive (practically at the same time) SNe natal kicks. A significant fraction of such newly formed BHs will escape from the system, except the most massive ones. This process substantially decreases the number of BHs in binaries and increases the number of single BHs kept in the system. Also, binaries that contain two massive stars and have small initial separations can merge during the common envelope evolution, producing a single BH instead of a BBH. Therefore, the number of single BHs will be much larger than the number following from the initial number of single stars, but much smaller than the number following from the initial total number of stars (single stars and binary components). Contrarily, for models with initially low binary fractions, there are fewer binaries to begin with and they are rather hard and will usually merge before the first SNe and form one massive BH with a small natal kick. Therefore, they will be kept in the system substantially increasing the number of formed (and kept) BHs from single stars. Accordingly, the number of BHs in binaries is very low - most BHs are single. According to equation 4 in \citet{BreenHeggie2013} the larger the BH mass fraction the larger the ratio of the half-mass radii for BHs and the whole system.    

\begin{figure}
\centering
\includegraphics[width=1.1\columnwidth]{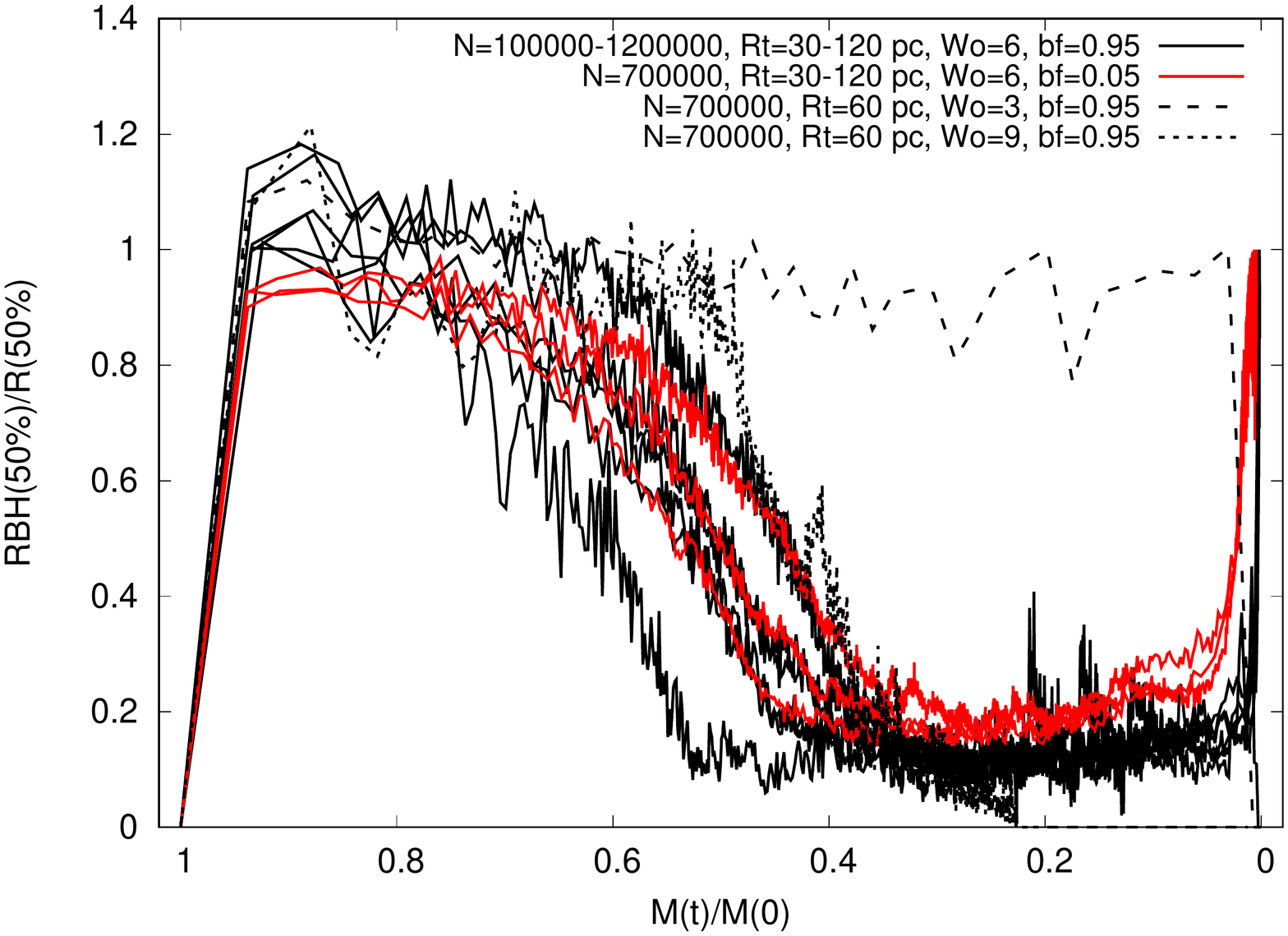}
\caption{Evolution of the ratio between BHS half-mass radius and overall half-mass radius as a function of the fraction of cluster bound mass for different \textsc{Mocca} models. Black lines - models with number of objects $N$ between $100000$ and $1200000$, $R_t$ between $30$ and $120$ pc, $W_0=6$ and binary fraction $0.95$. Red lines - models with number of objects $N=700000$, $R_t$ between $30$ and $120$ pc, $W_0=6$ and binary fraction $0.1$. Black dashed line - model with number of objects $N=700000$, $R_t=60$ pc, $W_0=3$ and binary fraction $0.95$.
Black doted line - model with number of objects $N=700000$, $R_t=60$ pc, $W_0=9$ and binary fraction $0.95$.}\label{RHBHSRH}
\end{figure}

Figure \ref{RHBHSRH} shows the evolution of the ratios between half-mass radii of BHS and the whole system, similar to Figure \ref{heggie}, but this time we are showing the half-mass radii from the \textsc{Mocca}  models and not the evaluated values from equation 4 in \citet{BreenHeggie2013}. Generally, both figures are very similar, after the initial phases of strong stellar evolution mass loss and BH mass segregation, which leads to the balanced cluster evolution. Models with lower numbers of objects achieve the phase of the balanced evolution for larger values of the fraction of cluster bound mass. This behaviour is connected with much faster BH mass segregation for smaller systems than for larger ones. The models with smaller binary fraction also show larger half-mass radii ratios as on Figure \ref{heggie}, but this time the difference is less pronounced. This suggests, as expected, that the structure of the BHSs in star clusters is more complicated than a simple two-component model. It also suggests, that the simple H{\'e}non principle \citep{Henon1975} properly captures the global evolution of star clusters, which have a complicated structure and contain a mixture of different kinds of objects responsible for the central energy generation. It is worth pointing out that the levels of the ratios of BHS half-mass radius and cluster half-mass radius are very similar on Figures \ref{heggie} and \ref{RHBHSRH}, which further strongly supports the theoretical picture of BHSs given by \citet{BreenHeggie2013}.

\begin{figure}
\centering
\includegraphics[width=1.1\columnwidth]{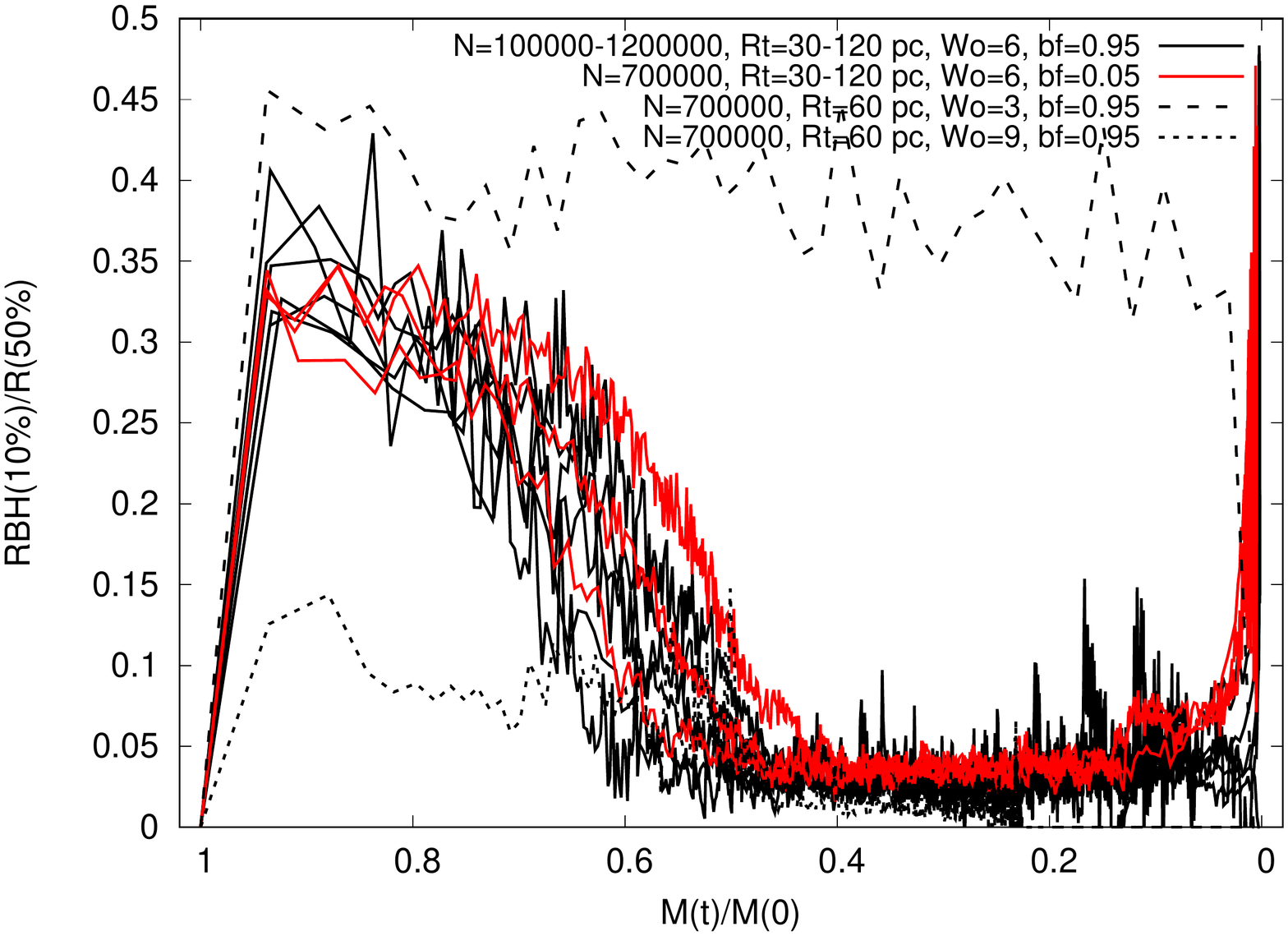}
\caption{Evolution of the ratio between $10\%$ Lagrangian radius for BHS and overall half-mass radius as a function of the fraction of cluster bound mass for different \textsc{Mocca} models. Black lines - models with number of objects $N$ between $100000$ and $1200000$, $R_t$ between $30$ and $120$ pc, $W_0=6$ and binary fraction $0.95$. Red lines - models with number of objects $N=700000$, $R_t$ between $30$ and $120$ pc, $W_0=6$ and binary fraction $0.1$. Black dashed line - model with number of objects $N=700000$, $R_t=60$ pc, $W_0=3$ and binary fraction $0.95$. Black doted line - model with number of objects $N=700000$, $R_t=60$ pc, $W_0=9$ and binary fraction $0.95$.}\label{R10BHSRH}
\end{figure}

The evolution of the ratio between $10\%$ BHS Lagrangian radius and cluster half-mass radius for different models is shown on Figure \ref{R10BHSRH}. Generally, the main features are very similar to Figure \ref{RHBHSRH}, but now the differences between models with low and large binary fractions are practically invisible. All models during balanced evolution have very similar ratios between $10\%$ BHS Lagrangian radius and cluster half-mass radius. This suggests that the energy generation by BHS is confined in the inner $10\%$ BHS Lagrangian radius and provides a universal path for cluster evolution regardless of the global cluster parameters.

\subsection{Cluster dissolution - Summary of the most important features}\label{summary}

In the previous Sections, we have provided arguments that tidally filling clusters harbouring BHSs in the centres can abruptly dissolve, provided that the clusters are described initially by a moderate King concentration parameter (around $W_0=6$), and cluster masses before the dissolution are about $10\%$ of the initial cluster masses. In this Section we will provide a short and coherent picture to take away from the paper.

\begin{figure}
\centering
\includegraphics[width=1.1\columnwidth]{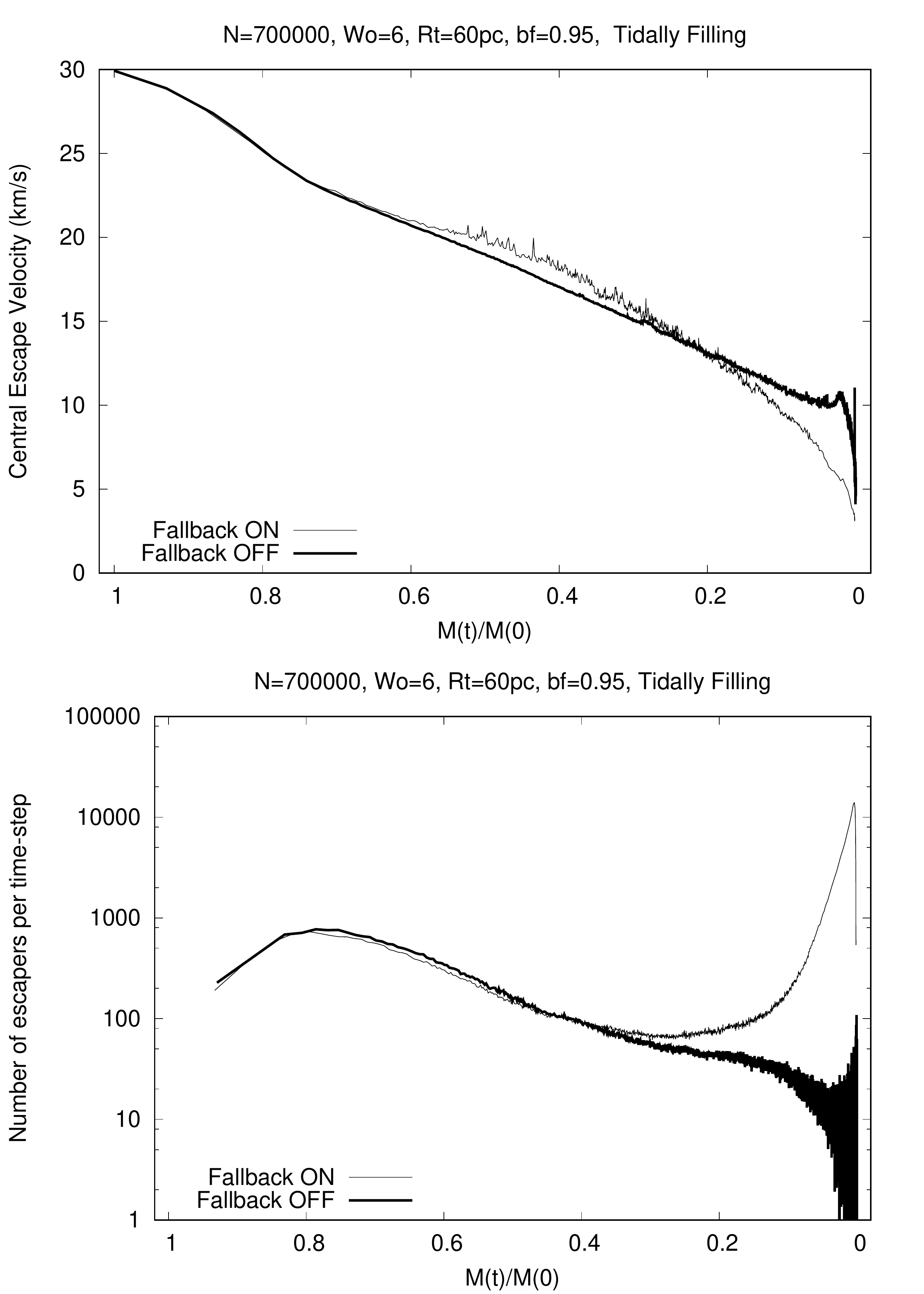}
\caption{Top panel: Evolution of the central escape velocity as a function of the fraction of cluster bound mass for the \textsc{Mocca} tidally filling model with mass fallback ON (black line) and mass fallback OFF (black dashed line), with 700000 objects (stars and binaries), $R_t=60$ pc, $W_0=6$, binary fraction $=0.95$. Bottom panel: Evolution of the escape rate as a function of the fraction of cluster bound mass for the same models as in the top panel. Escape rate is computed as a ratio between the number of all kind of escapers (relaxation, interactions, mass loss, tidal field) during a time-step divided by the duration of the time-step}\label{velo}
\end{figure}

The evolution of a tidally filling cluster is controlled by the following physical processes (if not stated otherwise all points below describe evolution of tidally filling models):
\begin{enumerate}
 \item initial mass loss is connected with stellar/binary evolution. All models regardless of their initial properties and for the canonical IMF \citep{Kroupa2001} are controlled by the mass loss up to the time of around 200 Myr and the fraction of cluster bound mass equal to about 0.7 - see Figures \ref{M-T-all} and \ref{M-T-N};
 \item then relaxation process takes over and BHs start to segregate and form a BHS. It seems that the BHS mass segregation stops around mass fraction equal to 0.45, depending on the number of stars, and takes slightly longer for the outer BHS Lagrangian radii. For tidally under-filling clusters, BH mass segregation is much faster and never stops for the outer BHS Lagrangian radii. This phase of evolution is characterized by: slightly decreasing average BH mass, constant BHS mass to cluster mass ratio and nearly constant Lagrangian radii for the BHS - see Figures \ref{ABH-M}, \ref{MBH-M} and \ref{R-M-tf} top panel, respectively. Massive BBHs generate enough energy to support cluster evolution without strongly changing BHS properties;
 \item a bit later, the fraction of cluster bound mass for a tidally filling cluster starts to differ from that for a tidally under-filling cluster - see Figure \ref{M-T-all}. This happens at mass fraction of about 0.3. The tidally under-filling cluster at this moment does not contain any BHS and energy is generated by stellar-type binaries. During this phase the tidal stripping becomes more and more efficient and the overall half-mass radius starts to slowly decrease, but BHS Lagrangian radii are still nearly constant. Accordingly, BHS mass slowly increases compared to the cluster mass and the average BH mass is slightly decreasing - see Figures \ref{R-M-tf}, \ref{MBH-M} and \ref{ABH-M}, respectively;
 \item at around mass fraction equal to 0.2, the cluster enters the phase of abrupt dissolution. During this phase, the average BH mass is nearly constant and the mass ratio between BHS mass and cluster mass starts to strongly increase. BHS Lagrangian radii start to slowly increase and the overall Lagrangian radii start to decrease with larger and larger speed. The virial ratio for BHS stops to decrease and becomes nearly constant, but the virial ratios for objects other than BHs start to increase showing enhanced removal of stars. The Cluster starts to evolve on a dynamical time-scale, not on a relaxation time-scale \citep{BanerjeeKroupa2011} - see Figures \ref{ABH-M}, \ref{MBH-M}, \ref{R-M-tf} top panel and \ref{Q-M}, respectively; 
 \item the strong energy generation by massive BHS, and as a consequence enhanced tidal stripping, are the main mechanisms responsible for abrupt cluster dissolution. The reasons are following:
\begin{enumerate}
\item as it was pointed out by \citet{ArcaSeddaetal2018} and \citet{Askaretal2018} BHs in BHS are very well mixed with other stars. The energy transport from BHS to the cluster halo occurs via stars kicked out from the BHS. Massive BBHs are extremely hard from the point of view of other, less massive, objects in the BHS, so during dynamical interactions, stars, and from time to time also BHs, are vigorously kicked out from the BHS;
\item some objects kicked out from the BHS will also escape from the cluster. The number of escaped objects will depend on the central escape velocity;
\item cluster is constantly tidally stripping, but the internal cluster structure is not strongly changing (see Figures \ref{R-M-tf} top panel and \ref{ABH-M}), so the escape velocity becomes smaller and smaller (see Figure \ref{velo} top panel) and more and more objects can be kicked out from the cluster - see Figure \ref{velo} bottom panel;
\item the positive feedback between strong energy generation, high flow of escaping stars, decreasing escape velocity and enhanced tidal stripping leads finally to loss of cluster dynamical equilibrium and abrupt dissolution - see Figure \ref{velo} bottom panel.
\end{enumerate}
 \end{enumerate} 

In our opinion, the above described third mechanism of cluster dissolution is universal and should work for all clusters, which harbour BHSs to the end of their life, and are exposed to a strong tidal field, regardless of the initial cluster structure or surrounding environment.

\section{Discussions and Conclusions}\label{disc}

In the previous Section we showed that tidally filling clusters with BHS in the center can show abrupt dissolution, provided that the cluster is described initially by a moderate King concentration parameter (around $W_0=6$), and the cluster mass before dissolution is about $10\%$ of the initial cluster mass. The dissolution is controlled by a strong energy generation in dynamical interactions inside the BHS and strong tidal stripping which lead to sudden loss of dynamical equilibrium. When the cluster loses a significant fraction of its initial mass, the BHS decouples from the rest of the cluster. The outer parts of the system cannot accommodate energy flux, because of increasing tidal stripping. Models that have larger $W_0$ or are strongly tidally under-filling are not able to keep their BHSs until the late phases of cluster evolution. They are characterized by slow dissolution controlled by the relaxation process and usually are core collapsed clusters. Tidally filling models with $W_0$ smaller than about $3$ will show abrupt dissolution which is caused by the strong early mass loss connected with stellar/binary evolution and extremely fast tidal stripping. They usually do not enter the core collapse phase.

\begin{figure}
\centering
\includegraphics[width=1.1\columnwidth]{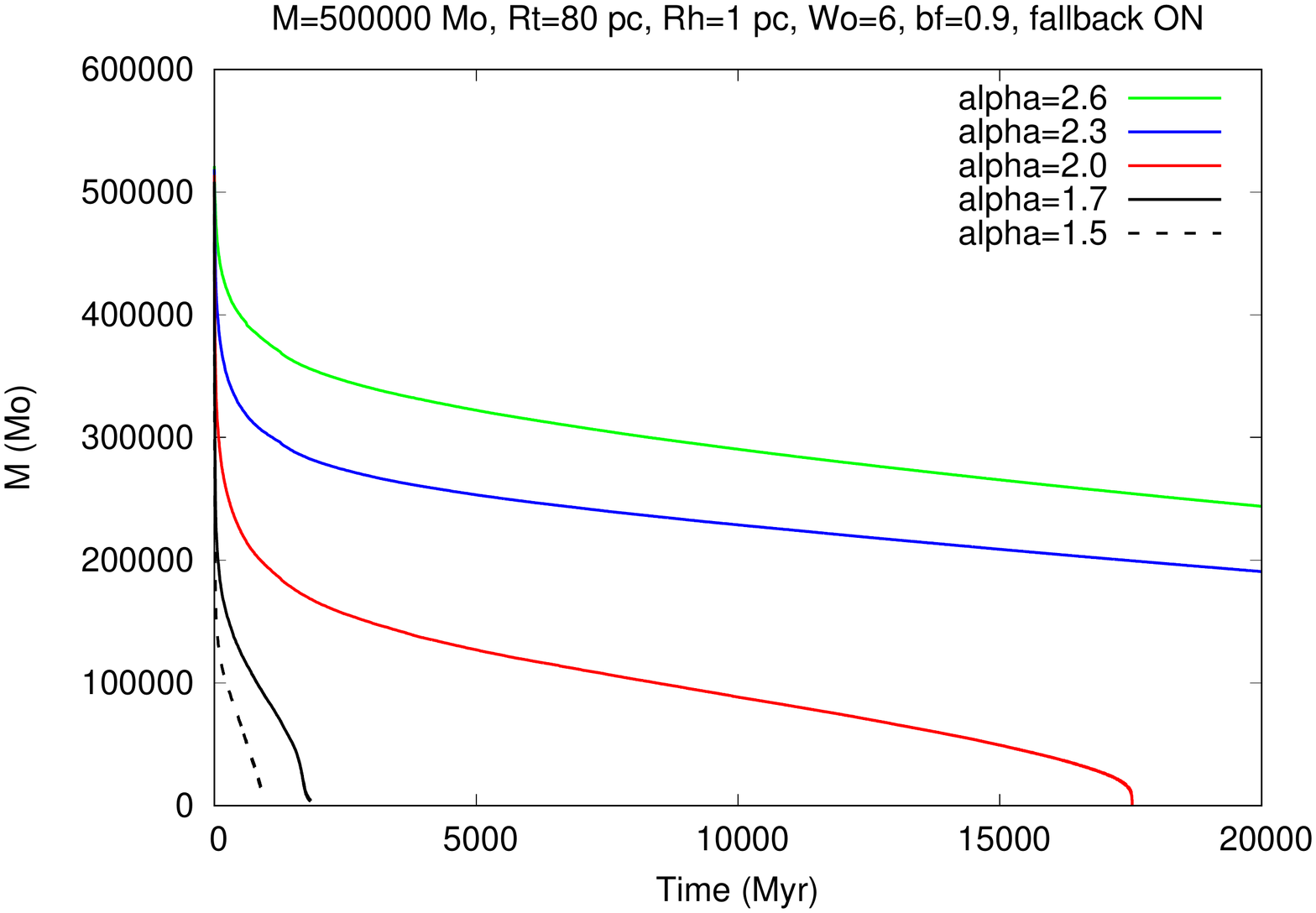}
\caption{Evolution of the cluster total mass as a function of time for \textsc{Mocca} models with different IMF power-low index for massive stars. Green line - $\alpha=2.6$. Blue line - $\alpha=2.3$. Red line - $\alpha=2.0$. Black line - $\alpha=1.7$. Black dashed line - $\alpha=1.5$.}\label{IMF}
\end{figure}

The above picture of the abrupt cluster dissolution by BHS energy generation and strong tidal field influence suggests that such a mechanism should operate not only in tidally filling clusters. For tidally under-filling clusters and/or clusters with larger $W_0$ we need initially a physical process which will quickly and strongly expand the cluster and/or decrease high initial concentration. Actually, such a process is operating in the initial phases of the cluster evolution, namely mass loss connected with the stellar/binary evolution. Only we need to make it more efficient. This can be achieved by making the IMF top-heavy. To check this hypothesis we started a new project to investigate how the cluster dissolution process depends on the IMF power-law index for massive stars \citep{Wangetal2019}. The very preliminary results are shown on Figure \ref{IMF} for a model with total mass equal to $500000$ $M_{\odot}$, $W_0=6$, strongly tidally under-filling with $R_t/R_h=80$, binary fraction equal to $90\%$, fallback ON and different IMF power-law index $\alpha$. As expected, models with top-heavy IMF show an abrupt dissolution which is very similar to the one described in this paper for tidally filling clusters with BHS. Models with extremely low $\alpha$ show dissolution which is powered by the initial mass loss connected with rapid stellar evolution not with energy generation by BHs. Very low $\alpha$ means a very large number of very massive stars which will very quickly lose most of their mass due to stellar evolution. Some indications about abrupt dissolution by top-heavy IMF were also provided by \citet{Chatterjeeetal2017}, but the authors did not discuss in detail the physical processes leading to such cluster dissolution. They only mentioned that strong energy generation could be responsible for such behaviour.

There are more and more observational evidences \citep[e.g.][]{Haghietal2017,Jerabkovaetal2018,Kalarietal2018,Schneideretal2018,
Hoseketal2019} and theory arguments \citep[e.g.][]{Marksetal2012} that star clusters are born with top-heavy IMF. As it was pointed out by \citet{Marksetal2012}, how strongly IMF is top-heavy depends on the environment properties (density, mass) of stellar forming regions. If globular clusters were formed in an environment with strong gas ram pressure, a stripping and quickly changing galactic tidal field \citep[e.g.][]{Kruijssen2014,Kruijssenetal2018,Renaud2018} we should expect that globular clusters that survive up to the Hubble time should be massive with large central concentrations. Probably clusters with smaller galactocentric distances were tidally filling or not strongly tidally under-filling. Cluster with larger distances, probably migrated out and should be strongly tidally under-filling \citep{Kruijssen2014}. Further spatial evolution of globular clusters will be shaped. among others, by two physical processes. One connected with dynamical friction, which drag clusters to the galactic center, and other connected with cluster dissolution, in some of them powered by BHSs. In both processes, globular clusters will be lost relatively early from the observational point of view. The latter process will lead to the buildup of the nuclear star cluster, removal of the most massive clusters and is the more effective one from the point of view of the number of retained BHs and the possible BH-BH mergers. The former process will relatively quickly disrupt clusters leading to a strong increase of BHs and BBHs populations in the galactic halo/bulge. Such a large population of BHs should be accompanied with some observational features connected with microlensing BHs \citep[e.g.][]{Rybickietal2018} and provide more firm constrains on the number of BHs floating in the Galactic halo.

According to the third mechanism of cluster dissolution the cluster dispersion is abrupt in time. The final stages of such a cluster should contain in the core a dark component surrounded by luminous stars. The dark core should measure about 1 pc and the cluster half-mass radius should be around 10 pc. The possibility of the formation of such 'dark clusters' in the population of open clusters has been predicted by \citet{BanerjeeKroupa2011}, but in this paper we extend this to the population of globular clusters. From the observational point of view, such clusters should be rather difficult to observe because they are very short living, contrary to the other type of 'dark clusters' proposed by \citet{Askar2017a}. Such 'dark clusters' are composed of massive intermediate mass BH surrounded by bright stars with a smaller total mass. Such clusters are characterized by much larger central velocity dispersion and can live in such a stage much longer than clusters described in this paper. It seems that the two types of 'dark clusters' proposed in the literature  have very different observational properties and should not be confused observationally.  

The \textsc{Mocca} code used to create the \textsc{Mocca Survey Database I} does not include post-Newtonian effects, which could be important during chaotic, strong dynamical interactions between BBHs and BHs in a dense and large BHSs. Because evolution of such systems is discussed in this paper, therefore we decided to comment on the possible influence of such effects on the cluster global structure and evolution. There are two main effects connected with gravitational wave emission: kinetic energy dissipation leading to an enhanced merging rate, and merger velocity kicks leading to an enhanced escape rate. Both effects will lead to the decrease of the BH number in the system. According to discussions in the previous Sections, it was argued that the most massive BBHs, composed of nearly equal mass BHs, are responsible for most of the energy generated in BHS and for slowing down the cluster evolution rate. If such a BBH will merge (because of gravitational wave radiation) they will be probably kicked out from the system due to higher recoil kicks for high mass ratio BBHs  \citep[see e.g.][]{Morawski2018}. Earlier removal of such a binary will lead to the formation of another, a bit less massive BBH, which in order to provide energy needed to support cluster structure will have to more frequently dynamically interact with other objects. This process, in turn will lead to the increase of cluster central density, faster dissolution of BHS and faster cluster evolution, preventing, in a most extreme situation, abrupt cluster dissolution. Taking into account strong stochasticity of dynamical interactions, we believe that such a scenario is rather unlikely. Firstly, most of the energy provided by massive BBH is generated with interactions with low mass ordinary stars, which are responsible for energy transport in the cluster. Such interactions will slowly increase the hardness of BBH, finally leading to its escape. Secondly, mergers between interacting BHs require very close approaches (comparable with the BH Schwarzschild radius), which in turn mean very hard BBH. It is possible that such hard BBH will be kicked out from the system before it can be prone for gravitational wave radiation mergers. Summarizing, we believe that the scenario of abrupt cluster dissolution discussed in this paper will be only slightly, if at all, affected by adding the post-Newtonian effects.

In this paper, we have presented on the basis of \textsc{Mocca} simulations of real star cluster evolution,  strong evidence of the existence of a third mechanism leading to cluster dissolution. In addition to cluster dissolution connected with strong mass loss due to stellar/binary evolution or connected with mass loss controlled by the relaxation process, we propose a dissolution mechanism connected with the presence of BHSs in tidally filling clusters. Energy generation in BHS together with strong tidal stripping leads to abrupt cluster dissolution. The dissolution process is connected with the loss of cluster dynamical equilibrium, similarly as it is in the case of dissolution powered by stellar evolution mass loss. We have argued that the third mechanism should be universal and will occur for star clusters embedded in a strong tidal field, which harbour massive BHS until the last phases of their evolution. Therefore, it should also work for a tidally under-filling cluster with top-heavy IMFs. This possibility will be further investigated in detail in the next paper.

\section*{Acknowledgements}

We wish to thank Douglas Heggie for the extremely insightful comments and in-depth discussions. MG and AL were partially supported by the Polish National Science Center (NCN) through the grant UMO-2016/23/B/ST9/02732. AA is currently supported by the Carl Tryggers Foundation for Scientific Research through the grant CTS 17:113 and was partially supported by NCN, Poland, through the grants UMO-2016/23/B/ST9/02732. LW in this work was supported by the funding from Alexander von Humboldt Foundation. AH was supported by Polish National Science Center grant 2016/20/S/ST9/00162. RS has been supported by the Alexander-von-Humboldt Polish Honorary Research Fellowship of the Foundation for Polish Science, and by the National Natural Science Foundation of China (NSFC), grant 11673032. We acknowledge support from the Chinese Academy of Sciences through the Silk Road Project at the National Astronomical Observatories, Chinese Academy of Sciences (NAOC), through the 'Qianren' special foreign experts programme of China, and through the Sino-German collaboration program GZ1284. This work benefited from support by the International Space Science Institute (ISSI), Bern, Switzerland, through its International Team programme ref. no. 393 The Evolution of Rich Stellar Populations \& BH Binaries (2017-18). 






\clearpage

\bibliographystyle{mnras}
\bibliography{references}

\begin{thebibliography}{}
\makeatletter
\relax
\def\mn@urlcharsother{\let\do\@makeother \do\$\do\&\do\#\do\^\do\_\do\%\do\~}
\def\mn@doi{\begingroup\mn@urlcharsother \@ifnextchar [ {\mn@doi@}
  {\mn@doi@[]}}
\def\mn@doi@[#1]#2{\def\@tempa{#1}\ifx\@tempa\@empty \href
  {http://dx.doi.org/#2} {doi:#2}\else \href {http://dx.doi.org/#2} {#1}\fi
  \endgroup}
\def\mn@eprint#1#2{\mn@eprint@#1:#2::\@nil}
\def\mn@eprint@arXiv#1{\href {http://arxiv.org/abs/#1} {{\tt arXiv:#1}}}
\def\mn@eprint@dblp#1{\href {http://dblp.uni-trier.de/rec/bibtex/#1.xml}
  {dblp:#1}}
\def\mn@eprint@#1:#2:#3:#4\@nil{\def\@tempa {#1}\def\@tempb {#2}\def\@tempc
  {#3}\ifx \@tempc \@empty \let \@tempc \@tempb \let \@tempb \@tempa \fi \ifx
  \@tempb \@empty \def\@tempb {arXiv}\fi \@ifundefined
  {mn@eprint@\@tempb}{\@tempb:\@tempc}{\expandafter \expandafter \csname
  mn@eprint@\@tempb\endcsname \expandafter{\@tempc}}}

\bibitem[\protect\citeauthoryear{{Aarseth}}{{Aarseth}}{2003}]{AarsethBOOK}
{Aarseth} S.~J.,  2003, {Gravitational N-Body Simulations}.
Cambridge University Press

\bibitem[\protect\citeauthoryear{{Ambartsumian}}{{Ambartsumian}}{1938}]{Ambartsumian1938}
{Ambartsumian} V.~A.,  1938, in English~translation: {Goodman} J.,  {Hut} P.,
  eds,  IAU Symposium Vol. 113, Dynamics of Star Clusters. Reidel, Dordrecht.
  pp 521, from Uch. Zap. L., G. U., 22, 19

\bibitem[\protect\citeauthoryear{{Arca Sedda}, {Askar}  \& {Giersz}}{{Arca
  Sedda} et~al.}{2018}]{ArcaSeddaetal2018}
{Arca Sedda} M.,  {Askar} A.,   {Giersz} M.,  2018, \mn@doi [\mnras]
  {10.1093/mnras/sty1859}, \href
  {http://adsabs.harvard.edu/abs/2018MNRAS.479.4652A} {479, 4652}

\bibitem[\protect\citeauthoryear{{Askar}, {Szkudlarek}, {Gondek-Rosi{\'n}ska},
  {Giersz}  \& {Bulik}}{{Askar} et~al.}{2017a}]{Askar2017b}
{Askar} A.,  {Szkudlarek} M.,  {Gondek-Rosi{\'n}ska} D.,  {Giersz} M.,
  {Bulik} T.,  2017a, \mn@doi [\mnras] {10.1093/mnrasl/slw177}, \href
  {http://adsabs.harvard.edu/abs/2017MNRAS.464L..36A} {464, L36}

\bibitem[\protect\citeauthoryear{{Askar}, {Bianchini}, {de Vita}, {Giersz},
  {Hypki}  \& {Kamann}}{{Askar} et~al.}{2017b}]{Askar2017a}
{Askar} A.,  {Bianchini} P.,  {de Vita} R.,  {Giersz} M.,  {Hypki} A.,
  {Kamann} S.,  2017b, \mn@doi [\mnras] {10.1093/mnras/stw2573}, \href
  {http://adsabs.harvard.edu/abs/2017MNRAS.464.3090A} {464, 3090}

\bibitem[\protect\citeauthoryear{{Askar}, {Arca Sedda}  \& {Giersz}}{{Askar}
  et~al.}{2018}]{Askaretal2018}
{Askar} A.,  {Arca Sedda} M.,   {Giersz} M.,  2018, \mn@doi [\mnras]
  {10.1093/mnras/sty1186}, \href
  {http://adsabs.harvard.edu/abs/2018MNRAS.478.1844A} {478, 1844}

\bibitem[\protect\citeauthoryear{{Banerjee} \& {Kroupa}}{{Banerjee} \&
  {Kroupa}}{2011}]{BanerjeeKroupa2011}
{Banerjee} S.,  {Kroupa} P.,  2011, \mn@doi [\apjl]
  {10.1088/2041-8205/741/1/L12}, \href
  {http://adsabs.harvard.edu/abs/2011ApJ...741L..12B} {741, L12}

\bibitem[\protect\citeauthoryear{{Baumgardt} \& {Makino}}{{Baumgardt} \&
  {Makino}}{2003}]{BaumgardtMakino2003}
{Baumgardt} H.,  {Makino} J.,  2003, \mn@doi [\mnras]
  {10.1046/j.1365-8711.2003.06286.x}, \href
  {http://adsabs.harvard.edu/abs/2003MNRAS.340..227B} {340, 227}

\bibitem[\protect\citeauthoryear{{Belczynski}, {Kalogera}  \&
  {Bulik}}{{Belczynski} et~al.}{2002}]{Belczynski2002}
{Belczynski} K.,  {Kalogera} V.,   {Bulik} T.,  2002, \mn@doi [\apj]
  {10.1086/340304}, \href {http://adsabs.harvard.edu/abs/2002ApJ...572..407B}
  {572, 407}

\bibitem[\protect\citeauthoryear{{Belloni}, {Zorotovic}, {Schreiber}, {Leigh},
  {Giersz}  \& {Askar}}{{Belloni} et~al.}{2017a}]{Belloni2017}
{Belloni} D.,  {Zorotovic} M.,  {Schreiber} M.~R.,  {Leigh} N.~W.~C.,  {Giersz}
  M.,   {Askar} A.,  2017a, \mn@doi [\mnras] {10.1093/mnras/stx575}, \href
  {http://ads.ari.uni-heidelberg.de/abs/2017MNRAS.468.2429B} {468, 2429}

\bibitem[\protect\citeauthoryear{{Belloni}, {Askar}, {Giersz}, {Kroupa}  \&
  {Rocha-Pinto}}{{Belloni} et~al.}{2017b}]{Belloni2017b}
{Belloni} D.,  {Askar} A.,  {Giersz} M.,  {Kroupa} P.,   {Rocha-Pinto} H.~J.,
  2017b, \mn@doi [\mnras] {10.1093/mnras/stx1763}, \href
  {http://adsabs.harvard.edu/abs/2017MNRAS.471.2812B} {471, 2812}

\bibitem[\protect\citeauthoryear{{Belloni}, {Kroupa}, {Rocha-Pinto}  \&
  {Giersz}}{{Belloni} et~al.}{2018}]{Belloni2018}
{Belloni} D.,  {Kroupa} P.,  {Rocha-Pinto} H.~J.,   {Giersz} M.,  2018, \mn@doi
  [\mnras] {10.1093/mnras/stx3034}, \href
  {http://adsabs.harvard.edu/abs/2018MNRAS.474.3740B} {474, 3740}

\bibitem[\protect\citeauthoryear{{Belloni}, {Giersz}, {Rivera Sandoval},
  {Askar}  \& {Cieciel{\aa}g}}{{Belloni} et~al.}{2019}]{Belloni2019}
{Belloni} D.,  {Giersz} M.,  {Rivera Sandoval} L.~E.,  {Askar} A.,
  {Cieciel{\aa}g} P.,  2019, \mn@doi [\mnras] {10.1093/mnras/sty3097}, \href
  {http://adsabs.harvard.edu/abs/2019MNRAS.483..315B} {483, 315}

\bibitem[\protect\citeauthoryear{{Breen} \& {Heggie}}{{Breen} \&
  {Heggie}}{2013}]{BreenHeggie2013}
{Breen} P.~G.,  {Heggie} D.~C.,  2013, \mn@doi [\mnras] {10.1093/mnras/stt628},
  \href {http://adsabs.harvard.edu/abs/2013MNRAS.432.2779B} {432, 2779}

\bibitem[\protect\citeauthoryear{{Chandrasekhar}}{{Chandrasekhar}}{1942}]{Chandrasekhar1942}
{Chandrasekhar} S.,  1942, {Principles of stellar dynamics}.
The University of Chicago press [1942]

\bibitem[\protect\citeauthoryear{{Chatterjee}, {Rodriguez}  \&
  {Rasio}}{{Chatterjee} et~al.}{2017}]{Chatterjeeetal2017}
{Chatterjee} S.,  {Rodriguez} C.~L.,   {Rasio} F.~A.,  2017, \mn@doi [\apj]
  {10.3847/1538-4357/834/1/68}, \href
  {http://adsabs.harvard.edu/abs/2017ApJ...834...68C} {834, 68}

\bibitem[\protect\citeauthoryear{{Chernoff} \& {Weinberg}}{{Chernoff} \&
  {Weinberg}}{1990}]{ChernoffWeinberg1990}
{Chernoff} D.~F.,  {Weinberg} M.~D.,  1990, \mn@doi [\apj] {10.1086/168451},
  \href {http://adsabs.harvard.edu/abs/1990ApJ...351..121C} {351, 121}

\bibitem[\protect\citeauthoryear{{Contenta}, {Varri}  \& {Heggie}}{{Contenta}
  et~al.}{2015}]{Contenta2015}
{Contenta} F.,  {Varri} A.~L.,   {Heggie} D.~C.,  2015, \mn@doi [\mnras]
  {10.1093/mnrasl/slv026}, \href
  {http://adsabs.harvard.edu/abs/2015MNRAS.449L.100C} {449, L100}

\bibitem[\protect\citeauthoryear{{Fregeau}, {Cheung}, {Portegies Zwart}  \&
  {Rasio}}{{Fregeau} et~al.}{2004}]{Fregeau2004}
{Fregeau} J.~M.,  {Cheung} P.,  {Portegies Zwart} S.~F.,   {Rasio} F.~A.,
  2004, \mn@doi [\mnras] {10.1111/j.1365-2966.2004.07914.x}, \href
  {http://adsabs.harvard.edu/abs/2004MNRAS.352....1F} {352, 1}

\bibitem[\protect\citeauthoryear{{Fukushige} \& {Heggie}}{{Fukushige} \&
  {Heggie}}{1995}]{FukushigeHeggie1995}
{Fukushige} T.,  {Heggie} D.~C.,  1995, \mn@doi [\mnras]
  {10.1093/mnras/276.1.206}, \href
  {http://adsabs.harvard.edu/abs/1995MNRAS.276..206F} {276, 206}

\bibitem[\protect\citeauthoryear{{Fukushige} \& {Heggie}}{{Fukushige} \&
  {Heggie}}{2000}]{Fukushige2000}
{Fukushige} T.,  {Heggie} D.~C.,  2000, \mn@doi [\mnras]
  {10.1046/j.1365-8711.2000.03811.x}, \href
  {http://adsabs.harvard.edu/abs/2000MNRAS.318..753F} {318, 753}

\bibitem[\protect\citeauthoryear{{Giersz}, {Heggie}  \& {Hurley}}{{Giersz}
  et~al.}{2008}]{Giersz2008}
{Giersz} M.,  {Heggie} D.~C.,   {Hurley} J.~R.,  2008, \mn@doi [\mnras]
  {10.1111/j.1365-2966.2008.13407.x}, \href
  {http://adsabs.harvard.edu/abs/2008MNRAS.388..429G} {388, 429}

\bibitem[\protect\citeauthoryear{{Giersz}, {Heggie}, {Hurley}  \&
  {Hypki}}{{Giersz} et~al.}{2013}]{Giersz2013}
{Giersz} M.,  {Heggie} D.~C.,  {Hurley} J.~R.,   {Hypki} A.,  2013, \mn@doi
  [\mnras] {10.1093/mnras/stt307}, \href
  {http://adsabs.harvard.edu/abs/2013MNRAS.431.2184G} {431, 2184}

\bibitem[\protect\citeauthoryear{{Giersz}, {Leigh}, {Hypki}, {L{\"u}tzgendorf}
  \& {Askar}}{{Giersz} et~al.}{2015}]{Giersz2015}
{Giersz} M.,  {Leigh} N.,  {Hypki} A.,  {L{\"u}tzgendorf} N.,   {Askar} A.,
  2015, \mn@doi [\mnras] {10.1093/mnras/stv2162}, \href
  {http://adsabs.harvard.edu/abs/2015MNRAS.454.3150G} {454, 3150}

\bibitem[\protect\citeauthoryear{{Gnedin} \& {Ostriker}}{{Gnedin} \&
  {Ostriker}}{1997}]{GnedinOstriker1997}
{Gnedin} O.~Y.,  {Ostriker} J.~P.,  1997, \mn@doi [\apj] {10.1086/303441},
  \href {http://adsabs.harvard.edu/abs/1997ApJ...474..223G} {474, 223}

\bibitem[\protect\citeauthoryear{{Haghi}, {Khalaj}, {Hasani Zonoozi}  \&
  {Kroupa}}{{Haghi} et~al.}{2017}]{Haghietal2017}
{Haghi} H.,  {Khalaj} P.,  {Hasani Zonoozi} A.,   {Kroupa} P.,  2017, \mn@doi
  [\apj] {10.3847/1538-4357/aa6719}, \href
  {http://adsabs.harvard.edu/abs/2017ApJ...839...60H} {839, 60}

\bibitem[\protect\citeauthoryear{{H{\'e}non}}{{H{\'e}non}}{1971}]{Henon1971}
{H{\'e}non} M.~H.,  1971, \mn@doi [\apss] {10.1007/BF00649201}, \href
  {http://adsabs.harvard.edu/abs/1971Ap%26SS..14..151H} {14, 151}

\bibitem[\protect\citeauthoryear{{H{\'e}non}}{{H{\'e}non}}{1975}]{Henon1975}
{H{\'e}non} M.,  1975, in {Hayli} A.,  ed.,  IAU Symposium Vol. 69, Dynamics of
  the Solar Systems. p.~133

\bibitem[\protect\citeauthoryear{{Hobbs}, {Lorimer}, {Lyne}  \&
  {Kramer}}{{Hobbs} et~al.}{2005}]{Hobbs2005}
{Hobbs} G.,  {Lorimer} D.~R.,  {Lyne} A.~G.,   {Kramer} M.,  2005, \mn@doi
  [\mnras] {10.1111/j.1365-2966.2005.09087.x}, \href
  {http://adsabs.harvard.edu/abs/2005MNRAS.360..974H} {360, 974}

\bibitem[\protect\citeauthoryear{{Hosek}, {Lu}, {Anderson}, {Najarro}, {Ghez},
  {Morris}, {Clarkson}  \& {Albers}}{{Hosek} et~al.}{2019}]{Hoseketal2019}
{Hosek} Jr. M.~W.,  {Lu} J.~R.,  {Anderson} J.,  {Najarro} F.,  {Ghez} A.~M.,
  {Morris} M.~R.,  {Clarkson} W.~I.,   {Albers} S.~M.,  2019, \mn@doi [\apj]
  {10.3847/1538-4357/aaef90}, \href
  {http://adsabs.harvard.edu/abs/2019ApJ...870...44H} {870, 44}

\bibitem[\protect\citeauthoryear{{Hurley}, {Pols}  \& {Tout}}{{Hurley}
  et~al.}{2000}]{Hurley2000}
{Hurley} J.~R.,  {Pols} O.~R.,   {Tout} C.~A.,  2000, \mn@doi [\mnras]
  {10.1046/j.1365-8711.2000.03426.x}, \href
  {http://adsabs.harvard.edu/abs/2000MNRAS.315..543H} {315, 543}

\bibitem[\protect\citeauthoryear{{Hurley}, {Tout}  \& {Pols}}{{Hurley}
  et~al.}{2002}]{Hurley2002}
{Hurley} J.~R.,  {Tout} C.~A.,   {Pols} O.~R.,  2002, \mn@doi [\mnras]
  {10.1046/j.1365-8711.2002.05038.x}, \href
  {http://adsabs.harvard.edu/abs/2002MNRAS.329..897H} {329, 897}

\bibitem[\protect\citeauthoryear{{Hypki} \& {Giersz}}{{Hypki} \&
  {Giersz}}{2013}]{HG2013}
{Hypki} A.,  {Giersz} M.,  2013, \mn@doi [\mnras] {10.1093/mnras/sts415}, \href
  {http://adsabs.harvard.edu/abs/2013MNRAS.429.1221H} {429, 1221}

\bibitem[\protect\citeauthoryear{{Jeans}}{{Jeans}}{1919}]{Jeans1919}
{Jeans} J.~H.,  1919, \mn@doi [\mnras] {10.1093/mnras/79.6.408}, \href
  {http://adsabs.harvard.edu/abs/1919MNRAS..79..408J} {79, 408}

\bibitem[\protect\citeauthoryear{{Je{\v r}{\'a}bkov{\'a}}, {Hasani Zonoozi},
  {Kroupa}, {Beccari}, {Yan}, {Vazdekis}  \& {Zhang}}{{Je{\v r}{\'a}bkov{\'a}}
  et~al.}{2018}]{Jerabkovaetal2018}
{Je{\v r}{\'a}bkov{\'a}} T.,  {Hasani Zonoozi} A.,  {Kroupa} P.,  {Beccari} G.,
   {Yan} Z.,  {Vazdekis} A.,   {Zhang} Z.-Y.,  2018, \mn@doi [\aap]
  {10.1051/0004-6361/201833055}, \href
  {http://adsabs.harvard.edu/abs/2018A%26A...620A..39J} {620, A39}

\bibitem[\protect\citeauthoryear{{Kalari}, {Carraro}, {Evans}  \&
  {Rubio}}{{Kalari} et~al.}{2018}]{Kalarietal2018}
{Kalari} V.~M.,  {Carraro} G.,  {Evans} C.~J.,   {Rubio} M.,  2018, \mn@doi
  [\apj] {10.3847/1538-4357/aab609}, \href
  {http://adsabs.harvard.edu/abs/2018ApJ...857..132K} {857, 132}

\bibitem[\protect\citeauthoryear{{King}}{{King}}{1966}]{King1966}
{King} I.~R.,  1966, \mn@doi [\aj] {10.1086/109857}, \href
  {http://adsabs.harvard.edu/abs/1966AJ.....71...64K} {71, 64}

\bibitem[\protect\citeauthoryear{{Kremer}, {Chatterjee}, {Ye}, {Rodriguez}  \&
  {Rasio}}{{Kremer} et~al.}{2019}]{Krameretal2019}
{Kremer} K.,  {Chatterjee} S.,  {Ye} C.~S.,  {Rodriguez} C.~L.,   {Rasio}
  F.~A.,  2019, \mn@doi [\apj] {10.3847/1538-4357/aaf646}, \href
  {http://ads.ari.uni-heidelberg.de/abs/2019ApJ...871...38K} {871, 38}

\bibitem[\protect\citeauthoryear{{Kroupa}}{{Kroupa}}{1995}]{Kroupa1995}
{Kroupa} P.,  1995, \mn@doi [\mnras] {10.1093/mnras/277.4.1491}, \href
  {http://adsabs.harvard.edu/abs/1995MNRAS.277.1491K} {277}

\bibitem[\protect\citeauthoryear{{Kroupa}}{{Kroupa}}{2001}]{Kroupa2001}
{Kroupa} P.,  2001, \mn@doi [\mnras] {10.1046/j.1365-8711.2001.04022.x}, \href
  {http://adsabs.harvard.edu/abs/2001MNRAS.322..231K} {322, 231}

\bibitem[\protect\citeauthoryear{{Kroupa}}{{Kroupa}}{2011}]{Kroupa2011}
{Kroupa} P.,  2011, in {Alves} J.,  {Elmegreen} B.~G.,  {Girart} J.~M.,
  {Trimble} V.,  eds,  IAU Symposium Vol. 270, Computational Star Formation. pp
  141--149 (\mn@eprint {arXiv} {1012.1596}), \mn@doi{10.1017/S1743921311000305}

\bibitem[\protect\citeauthoryear{{Kruijssen}}{{Kruijssen}}{2014}]{Kruijssen2014}
{Kruijssen} J.~M.~D.,  2014, \mn@doi [Classical and Quantum Gravity]
  {10.1088/0264-9381/31/24/244006}, \href
  {http://adsabs.harvard.edu/abs/2014CQGra..31x4006K} {31, 244006}

\bibitem[\protect\citeauthoryear{{Kruijssen}, {Pfeffer}, {Reina-Campos},
  {Crain}  \& {Bastian}}{{Kruijssen} et~al.}{2018}]{Kruijssenetal2018}
{Kruijssen} J.~M.~D.,  {Pfeffer} J.~L.,  {Reina-Campos} M.,  {Crain} R.~A.,
  {Bastian} N.,  2018, \mn@doi [\mnras] {10.1093/mnras/sty1609}, \href
  {http://adsabs.harvard.edu/abs/2018MNRAS.tmp.1537K} {}

\bibitem[\protect\citeauthoryear{{Leigh}, {Giersz}, {Marks}, {Webb}, {Hypki},
  {Heinke}, {Kroupa}  \& {Sills}}{{Leigh} et~al.}{2015}]{Leigh2015}
{Leigh} N.~W.~C.,  {Giersz} M.,  {Marks} M.,  {Webb} J.~J.,  {Hypki} A.,
  {Heinke} C.~O.,  {Kroupa} P.,   {Sills} A.,  2015, \mn@doi [\mnras]
  {10.1093/mnras/stu2110}, \href
  {http://adsabs.harvard.edu/abs/2015MNRAS.446..226L} {446, 226}

\bibitem[\protect\citeauthoryear{{Marks}, {Kroupa}, {Dabringhausen}  \&
  {Pawlowski}}{{Marks} et~al.}{2012}]{Marksetal2012}
{Marks} M.,  {Kroupa} P.,  {Dabringhausen} J.,   {Pawlowski} M.~S.,  2012,
  \mn@doi [\mnras] {10.1111/j.1365-2966.2012.20767.x}, \href
  {http://adsabs.harvard.edu/abs/2012MNRAS.422.2246M} {422, 2246}

\bibitem[\protect\citeauthoryear{{Milone} et~al.,}{{Milone}
  et~al.}{2012}]{Milone2012}
{Milone} A.~P.,  et~al., 2012, \mn@doi [\aap] {10.1051/0004-6361/201016384},
  \href {http://adsabs.harvard.edu/abs/2012A%26A...540A..16M} {540, A16}

\bibitem[\protect\citeauthoryear{{Morawski}, {Giersz}, {Askar}  \&
  {Belczynski}}{{Morawski} et~al.}{2018}]{Morawski2018}
{Morawski} J.,  {Giersz} M.,  {Askar} A.,   {Belczynski} K.,  2018, \mn@doi
  [\mnras] {10.1093/mnras/sty2401}, \href
  {http://adsabs.harvard.edu/abs/2018MNRAS.481.2168M} {481, 2168}

\bibitem[\protect\citeauthoryear{{Renaud}}{{Renaud}}{2018}]{Renaud2018}
{Renaud} F.,  2018, \mn@doi [\nar] {10.1016/j.newar.2018.03.001}, \href
  {http://adsabs.harvard.edu/abs/2018NewAR..81....1R} {81, 1}

\bibitem[\protect\citeauthoryear{{Rybicki}, {Wyrzykowski}, {Klencki}, {de
  Bruijne}, {Belczy{\'n}ski}  \& {Chru{\'s}li{\'n}ska}}{{Rybicki}
  et~al.}{2018}]{Rybickietal2018}
{Rybicki} K.~A.,  {Wyrzykowski} {\L}.,  {Klencki} J.,  {de Bruijne} J.,
  {Belczy{\'n}ski} K.,   {Chru{\'s}li{\'n}ska} M.,  2018, \mn@doi [\mnras]
  {10.1093/mnras/sty356}, \href
  {http://adsabs.harvard.edu/abs/2018MNRAS.476.2013R} {476, 2013}

\bibitem[\protect\citeauthoryear{{Schneider} et~al.,}{{Schneider}
  et~al.}{2018}]{Schneideretal2018}
{Schneider} F.~R.~N.,  et~al., 2018, \mn@doi [Science]
  {10.1126/science.aan0106}, \href
  {http://adsabs.harvard.edu/abs/2018Sci...359...69S} {359, 69}

\bibitem[\protect\citeauthoryear{{Spitzer}}{{Spitzer}}{1987}]{Spitzer1987}
{Spitzer} L.,  1987, {Dynamical evolution of globular clusters}.
Princeton University Press

\bibitem[\protect\citeauthoryear{{Stodolkiewicz}}{{Stodolkiewicz}}{1982}]{Stod1982}
{Stodolkiewicz} J.~S.,  1982, \actaa, \href
  {http://adsabs.harvard.edu/abs/1982AcA....32...63S} {32, 63}

\bibitem[\protect\citeauthoryear{{Stodolkiewicz}}{{Stodolkiewicz}}{1986}]{Stod1986}
{Stodolkiewicz} J.~S.,  1986, \actaa, \href
  {http://adsabs.harvard.edu/abs/1986AcA....36...19S} {36, 19}

\bibitem[\protect\citeauthoryear{{Tanikawa} \& {Fukushige}}{{Tanikawa} \&
  {Fukushige}}{2009}]{TanikawaFukushige2009}
{Tanikawa} A.,  {Fukushige} T.,  2009, \mn@doi [\pasj] {10.1093/pasj/61.4.721},
  \href {http://adsabs.harvard.edu/abs/2009PASJ...61..721T} {61, 721}

\bibitem[\protect\citeauthoryear{{Vink}, {de Koter}  \& {Lamers}}{{Vink}
  et~al.}{2001}]{Vink2001}
{Vink} J.~S.,  {de Koter} A.,   {Lamers} H.~J.~G.~L.~M.,  2001, \mn@doi [\aap]
  {10.1051/0004-6361:20010127}, \href
  {http://adsabs.harvard.edu/abs/2001A%26A...369..574V} {369, 574}

\bibitem[\protect\citeauthoryear{{Wang}, {Spurzem}, {Aarseth}, {Nitadori},
  {Berczik}, {Kouwenhoven}  \& {Naab}}{{Wang} et~al.}{2015}]{Wang2015}
{Wang} L.,  {Spurzem} R.,  {Aarseth} S.,  {Nitadori} K.,  {Berczik} P.,
  {Kouwenhoven} M.~B.~N.,   {Naab} T.,  2015, \mn@doi [\mnras]
  {10.1093/mnras/stv817}, \href
  {http://adsabs.harvard.edu/abs/2015MNRAS.450.4070W} {450, 4070}

\bibitem[\protect\citeauthoryear{{Wang}, {Giersz}  \& {Askar}}{{Wang}
  et~al.}{2019}]{Wangetal2019}
{Wang} L.,  {Giersz} M.,   {Askar} A.,  2019, in preparation

\bibitem[\protect\citeauthoryear{{Weinberg}}{{Weinberg}}{1993}]{Weinberg1993}
{Weinberg} M.~D.,  1993, in {Smith} G.~H.,  {Brodie} J.~P.,  eds,  Astronomical
  Society of the Pacific Conference Series Vol. 48, The Globular Cluster-Galaxy
  Connection. p.~689

\bibitem[\protect\citeauthoryear{{Whitehead}, {McMillan}, {Vesperini}  \&
  {Portegies Zwart}}{{Whitehead} et~al.}{2013}]{Whitehead2013}
{Whitehead} A.~J.,  {McMillan} S.~L.~W.,  {Vesperini} E.,   {Portegies Zwart}
  S.,  2013, \mn@doi [\apj] {10.1088/0004-637X/778/2/118}, \href
  {http://adsabs.harvard.edu/abs/2013ApJ...778..118W} {778, 118}

\makeatother
\end{thebibliography}

\end{document}